\documentclass[%
	aps, pra,
	twocolumn, superscriptaddress,
	10pt,
	floatfix, nobibnotes,
	tightenlines,
	longbibliography
]{revtex4-1}

\usepackage[utf8]{inputenc}
\usepackage{lmodern}
\usepackage{amssymb,amsmath,amsthm}
\usepackage{mathtools}
\usepackage{amsbsy}
\usepackage{bm,bbm}
\usepackage[english]{babel}

\usepackage[dvipsnames]{xcolor}
\usepackage{color,graphicx}
\graphicspath{{./figures/}}

\usepackage{physics}

\usepackage{epsfig}

\usepackage[colorlinks]{hyperref}
\usepackage[capitalize]{cleveref}
\crefname{pluralequation}{eqs.}{eqs.}
\Crefname{pluralequation}{Eqs.}{Eqs.}

\usepackage{subfigure}

\usepackage{microtype}

\usepackage{soulutf8}

\usepackage{xstring}

\usepackage{acronym}
\newacro{DTQW}{Discrete Time Quantum Walk}
\newacro{OAM}{Orbital Angular Momentum}
\newacro{SAM}{Spin Angular Momentum}

\usepackage{todonotes}
\presetkeys{todonotes}{size=\tiny}{}

\usepackage{booktabs}

\newcommand{\dimenst}[2]{#1$\times$#2}

\newcommand{\bs}[1]{\boldsymbol{#1}}

\newcommand\scalemath[2]{\scalebox{#1}{\mbox{\ensuremath{\displaystyle #2}}}}

\begin{document}


\title{Quantum state engineering using one-dimensional discrete-time quantum walks}

\author{Luca Innocenti}
\affiliation{Centre for Theoretical Atomic, Molecular, and Optical Physics, School of Mathematics and Physics, Queen's University Belfast, BT7 1NN Belfast, United Kingdom}
\author{Helena Majury} 
\affiliation{Centre for Secure Information Technologies, Queen's University, Belfast BT7 1NN Belfast, United Kingdom}
\affiliation{Centre for Theoretical Atomic, Molecular, and Optical Physics, School of Mathematics and Physics, Queen's University Belfast, BT7 1NN Belfast, United Kingdom}
\author{Taira Giordani}
\affiliation{Dipartimento di Fisica, Sapienza Universit\`{a} di Roma,
Piazzale Aldo Moro 5, I-00185 Roma, Italy} 
\author{Nicol\`o Spagnolo}
\affiliation{Dipartimento di Fisica, Sapienza Universit\`{a} di Roma,
Piazzale Aldo Moro 5, I-00185 Roma, Italy}
\author{Fabio Sciarrino}
\affiliation{Dipartimento di Fisica, Sapienza Universit\`{a} di Roma,
Piazzale Aldo Moro 5, I-00185 Roma, Italy}
\affiliation{Consiglio Nazionale delle Ricerche, Istituto dei sistemi Complessi (CNR-ISC), Via dei Taurini 19, 00185 Roma, Italy}
\author{Mauro Paternostro}
\affiliation{Centre for Theoretical Atomic, Molecular, and Optical Physics, School of Mathematics and Physics, Queen's University Belfast, BT7 1NN Belfast, United Kingdom}
\affiliation{Laboratoire Kastler Brossel, ENS-PSL Research University, 24 rue Lhomond, F-75005 Paris, France}
\author{Alessandro Ferraro}
\affiliation{Centre for Theoretical Atomic, Molecular, and Optical Physics, School of Mathematics and Physics, Queen's University Belfast, BT7 1NN Belfast, United Kingdom}

\date{\today}

\begin{abstract}
Quantum state preparation in high-dimensional systems is an essential requirement for many quantum-technology applications. The engineering of an arbitrary quantum state is, however, typically strongly dependent on the experimental platform chosen for implementation, and a general framework is still missing.
Here we show that coined quantum walks on a line, which represent a framework general enough to encompass a variety of different platforms, can be used for quantum state engineering of arbitrary superpositions of the walker's sites.
We achieve this goal by identifying a set of conditions that fully characterize the reachable states in the space comprising walker and coin,
and providing a method to efficiently compute the corresponding set of coin parameters.
We assess the feasibility of our proposal by identifying a linear optics experiment based on photonic orbital angular momentum technology.
\end{abstract}
\maketitle

\section{Introduction}
Quantum states of high-dimensional Hilbert spaces are of paramount interest both from a foundational and applicative perspective. They exhibit a richer entanglement structure \cite{horodecki2009quantum} and a stronger violation of local realism \cite{brunner2014bell} than their qubit counterpart. Even at the level of a single system, they illustrate the contextual character of quantum mechanics in a way that cannot result from entanglement \cite{lapkiewicz2011experimental}. 
In the framework of quantum communication, high-dimensional systems guarantee higher security and increased transmission rates \cite{bechmannpasquinucci2000quantum, cerf2002security, bru2002optimal, acin2003security, karimipour2002quantum, durt2004security, nunn2013largealphabet, mower2013highdimensional, lee2014entanglementbased, zhong2015photonefficient}, allowing also for convenient solutions to problems such as quantum bit commitment \cite{langford2004measuring} and  Byzantine agreement \cite{fitzi2001quantum}.
More generally, they have been shown to be advantageous for various applications,
from spatial imaging \cite{howland2013efficient} to quantum computation \cite{bartlett2002quantum, ralph2007efficient} and error correction \cite{campbell2012magicstate}.

Not surprisingly, in the past decades there has been a steady interest in engineering quantum states of high dimensions. 
Many engineering strategies have been theoretically proposed and experimentally realised in quantum optical systems using a variety of degrees of freedom, from time-energy \cite{thew2004belltype, bessire2014versatile} to polarization \cite{bogdanov2004qutrit}, path \cite{osullivanhale2005pixel}, orbital angular momentum \cite{mair2001entanglement, mclaren2012entangled, krenn2013entangled, krenn2014generation, zhang2016engineering} and frequency \cite{bernhard2013shaping, jin2016simple}. In general, such strategies depend strongly on the specific setting under consideration and a unified framework is lacking. 

Here we make a significant step in this direction by proposing a strategy based on the ubiquitous dynamics offered by quantum walks, which are quantum generalizations of classical random walks~\cite{aharonov1993quantum, nayak2000quantum, ambainis2001onedimensional, kempe2003quantum, venegasandraca2012quantum}.
In its simplest form, a quantum walk involves a high-dimensional system (generically a $d$-dimensional system dubbed \textit{qudit}), usually referred to as \emph{walker}, endowed with an inner 2-dimensional degree of freedom, referred to as the \emph{coin state} of the walker. 
At every step, the coin state is \emph{flipped} with some unitary operation, and the walker moves coherently left and right, conditionally on the coin state.
We focus on \acp{DTQW} on a line~\cite{ambainis2001onedimensional}, allowing the coin operation to change from step to step, while remaining site-independent~\cite{ribeiro2004aperiodic,wjcik2004quasiperiodic,bauls2006quantum}.
We demonstrate the effectiveness of this framework for the state engineering of $d$-dimensional systems and provide a set of efficiently verifiable necessary and sufficient conditions for a given state to be the output of a quantum walk evolution.	

Our results provide an additional relevant instance of the richness of quantum walks for quantum information processing tasks, notwithstanding their simplicity.
As other notable examples, it is worth mentioning that both the DTQW~\cite{aharonov1993quantum} and its continuous-time variant~\cite{farhi1998quantum} are universal for quantum computation~\cite{childs2009universal,childs2013universal},
and allow for efficient implementations of quantum search algorithms~%
\cite{shenvi2003quantum,ambainis2005coins,tulsi2008faster}.
This has led to several experimental implementations with a variety of architectures~%
\cite{ct2006quantum, schwartz2007transport, chandrashekar2008quantum, perets2008realization, karski2009quantum, schmitz2009quantum, zhringer2010realization, peruzzo2010quantum, owens2011twophoton, weitenberg2011singlespin, giuseppe2013einsteinpodolskyrosen, fukuhara2013microscopic, poulios2014quantum, preiss2015strongly, chapman2016experimental, caruso2016fast}.
In particular, discrete-time quantum walks have been demonstrated in a variety of photonic platforms,
including linear optical interferometers~\cite{sansoni2012twoparticle,crespi2013anderson,harris2015bosonic,pitsios2016photonic},
intrinsically stable multi-mode interferometers with polarization optics~\cite{broome2010discrete,kitagawa2012observation,vitelli2013joining},
fiber-loop systems for time-bin encoding~\cite{schreiber2010photons,schreiber2012a,boutari2016large},
and quantum walks in the orbital angular momentum space~\cite{cardano2015quantum,cardano2016statistical}.
We take advantage of the suitability of photonic settings for the implementation of DTQW to present an experimental proposal for the linear-optics validation of our results, exploiting both the polarization and orbital angular momentum degrees of freedom of a photon.

\section{Quantum Walks background}
\begin{figure}[ht!]
\includegraphics[width=0.59\columnwidth]{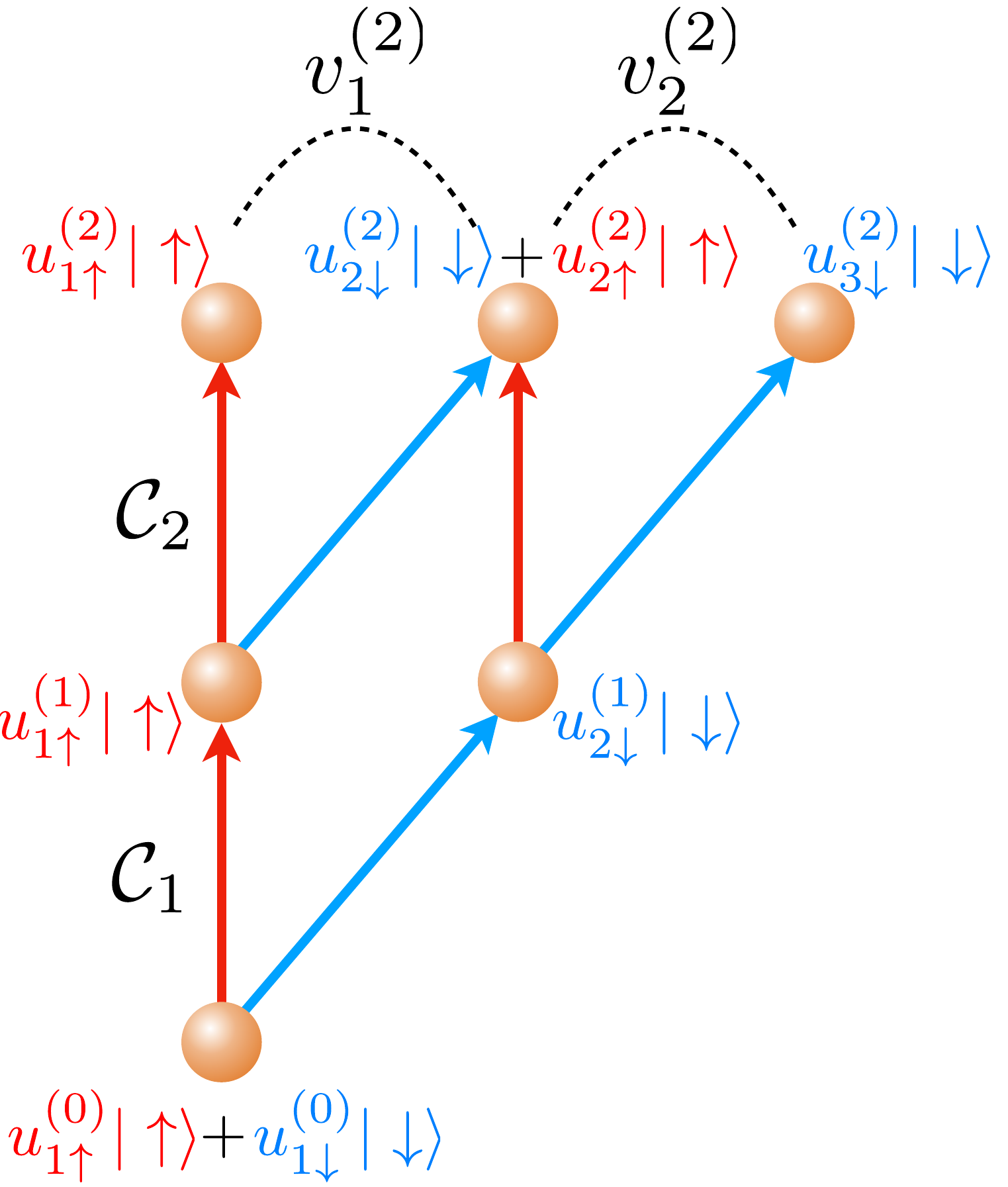}
	\caption{
		Schematic representation of how the amplitudes are distributed among the sites, at the various steps of the evolution.
		The red (blue) arrows represent the movement of the walker with coin state $\ket\uparrow$ ($\ket\downarrow$) after the coin flip.
		The coin operators $\mathcal C_i$ determine the weights with which the amplitude at every site is distributed between the red and blue arrows, and thus the two sites of the next layer with which it is connected.
	}
	\label{fig:walker}
\end{figure}

A single step of quantum walk evolution consists of a \emph{coin flipping} step, during which a unitary { transformation} $\mathcal C$ is applied to the coin, and a \emph{walking step}, in which the walker's state evolves conditionally to the state of the coin, through a controlled-shift operator $\mathcal S$.
Formally, given an initial state of the walker
$
	\ket\Psi \equiv \sum_{k=1}^n \sum_{s\in\{\uparrow,\downarrow\}}
	u_{k,s} \ket{k} \otimes \ket{s},
$
after one step the state evolves to
\begin{equation}
	\mathcal W_{\mathcal C} \ket\Psi \equiv
	\mathcal S \,\mathcal C \ket\Psi
	= \sum_{k=1}^n \sum_{s\in\{\uparrow,\downarrow\}}
	u_{k,s}
	\mathcal S \big( \ket{k} \otimes \mathcal C \ket{s} \big),
	\label{eq:step_operator_definition}
\end{equation}
where we defined the \emph{step operator} $\mathcal W_{\mathcal C}$ as the combined action of a coin flip and a controlled shift, and $\mathcal S$ is given by
\begin{equation}
	\mathcal S =
	\sum_k \big(
		\ketbra{k}{k} \otimes \ketbra{\uparrow}{\uparrow}
		+
		\ketbra{k + 1}{k} \otimes \ketbra{\downarrow}{\downarrow}
	\big).
	\label{eq:controlled_shift_operator}
\end{equation}
We here assume the walker's Hilbert space to be infinite dimensional (or, equivalently, larger than the considered number of steps), so that there is no need to take the boundary conditions into account.
It is worth noting that we use a slightly different convention than those commonly found in the literature. Rather than considering the walker to be moving left or right conditionally to different coin states, we assume the walker to stand still or move right depending on whether the coin is prepared in $\ket{\uparrow}$ or $\ket{\downarrow}$ (cf.~\cref{fig:walker})~\cite{hoyer2009faster, montero2013unidirectional,montero2015quantum}.
While for 2-dimensional coin states the two arrangements are equivalent, our choice allows for both a clearer presentation and more efficient simulations.

\section{Reachability conditions}
\label{sec:reachability}
\subsection{1-step reachability condition}
\label{sec:1step_reachability}
Let us now focus on the case where the walker is initially localised at site $i=1$ with some unspecified coin state, so that the initial state of the system reads
${\ket*{\Psi^{(0)}} = \ket{1}\otimes(u^{(0)}_{1,\uparrow} \ket{1,\uparrow} + u^{(0)}_{1,\downarrow} \ket{1,\downarrow}})$.
Given the definition of the conditional shift operation $\mathcal S$, after a step with coin operator $\mathcal C_1$, the resulting state ${\ket*{\Psi^{(1)}} \equiv \mathcal W_{\mathcal C_1} \ket*{\Psi^{(0)}}}$ satisfies the conditions 
${\braket{1, \downarrow}{\Psi^{(1)}} = \braket{2, \uparrow}{\Psi^{(1)}} = 0}$.
More generally, for any initial walker state $\ket{\Psi}$ spanning $n$ sites, the application of $\mathcal W_{\mathcal C_1}$ produces a state spanning $n+1$ sites and satisfying the equations
\begin{equation}
	\mel{1, \downarrow}{\mathcal W_{\mathcal C_1}}{\Psi} =
	\mel{n+1, \uparrow}{\mathcal W_{\mathcal C_1}}{\Psi} = 0.
	\label{eq:vanishing_endpoints}
\end{equation}
The implication goes both ways: any state $\ket\Phi$ of a system spanning $n+1$ sites and with an additional spin-like degree of freedom, such that
$
\braket{1, \downarrow}{\Phi} =
\braket{n+1, \uparrow}{\Phi} = 0
$,
is the output of one step of a quantum walk evolution with suitable coin operator and initial coin state.

\subsection{2-step reachability conditions}
\label{sec:2step_reachability}
Consider now another coin operator $\mathcal C_2$, and the state
$\ket*{\Psi^{(2)}} \equiv \mathcal W_{\mathcal C_2} \ket*{\Psi^{(1)}}$.
It is useful to describe the evolution of the amplitudes $u_{i, s}^{(n)}$ after a generic step $\mathcal W_{\mathcal C}$ as
\begin{equation}
	\bs{v}_i^{(n+1)} \equiv
	\begin{pmatrix}
		u^{(n+1)}_{i, \uparrow} \\ u^{(n+1)}_{i+1, \downarrow}
	\end{pmatrix}
	= \mathcal C
	\begin{pmatrix}
		u^{(n)}_{i, \uparrow} \\ u^{(n)}_{i, \downarrow}
	\end{pmatrix},
	\label{eq:evolution_equation_with_C}
\end{equation}
where the complex vectors $\bs v_i^{(n)}$ collect the amplitudes at the $n$-th step that came from the $i$-th site at the previous step.
It is easily verified that this description is equivalent to \cref{eq:step_operator_definition}.
Applying \cref{eq:evolution_equation_with_C} to $\ket*{\Psi^{(1)}}$ we get
\begin{equation}
	\bs{v}^{(2)}_1
	=
	\mathcal C_2
	\begin{pmatrix}
		u^{(1)}_{1,\uparrow} \\ 0
	\end{pmatrix},
	\quad
	\bs{v}^{(2)}_2
	=
	\mathcal C_2
	\begin{pmatrix}
		0 \\ u^{(1)}_{2,\downarrow} 
	\end{pmatrix},
	\label{eq:leading_to_first_orthogonality}
\end{equation}
which directly implies the orthogonality of
$\bs v_1^{(2)}$ and $\bs v_{2}^{(2)}$,
for any unitary $\mathcal C_2$.
More explicitly, the amplitudes of $\ket*{\Psi^{(2)}}$ satisfy the conditions
\begin{subequations}
	\begin{align}
		u^{(2)}_{1, \downarrow} = u^{(2)}_{3, \uparrow} = 0,
		\label{eq:vanishing_endpoints_after_2steps}\\
		u^{(2)*}_{1, \uparrow} u^{(2)}_{2, \uparrow}
		+ u^{(2)*}_{2, \downarrow} u^{(2)}_{3, \downarrow} = 0.
		\label{eq:1st_orthogonality_condition}
	\end{align}
	\label[pluralequation]{eq:both_2ndstep_conditions}
\end{subequations}
On the other hand, if $\ket\Phi$ is a walker state spanning $3$ sites like $\ket*{\Psi^{(2)}}$, with amplitudes satisfying the above relations,
then \cref{eq:vanishing_endpoints_after_2steps} ensures that $\ket\Phi$ is output of at least 1 step,
while \cref{eq:1st_orthogonality_condition} implies the existence of a unitary operator $\mathcal C_2$ and complex numbers $a$ and $b$ such that
\begin{equation}
	\begin{pmatrix}
		u_{1, \uparrow} \\ u_{2, \downarrow}
	\end{pmatrix}
	=
	\mathcal C_2
	\begin{pmatrix}
		a \\ 0
	\end{pmatrix},
	\qquad
	\begin{pmatrix}
		u_{2, \uparrow} \\ u_{3, \downarrow}
	\end{pmatrix}
	=
	\mathcal C_2
	\begin{pmatrix}
		0 \\ b
	\end{pmatrix}.
	\label{eq:from_us_to_coin_operator}
\end{equation}
Note that, for fixed values of the final amplitudes $u_{i,s}$, the coin operator $\mathcal C_2$ (and consequently the amplitudes $a$ and $b$) is determined up to two phases.
In other words, for any $\mathcal C_2$ satisfying \cref{eq:from_us_to_coin_operator}, the matrix
$\mathcal C_2 \begin{pmatrix}
	e^{i\theta_1} & 0 \\ 0 & e^{i\theta_2}
\end{pmatrix}$
is another possible coin operator generating $\ket\Phi$, corresponding to the extremal amplitudes at the previous step $e^{-i\theta_1}a$ and $e^{-i\theta_2}b$.
Given that quantum states are always defined up to a global phase, this implies that the only freedom in the choice of the coin operator is in the phase difference between the two columns.
A step backwards with coin operator $\mathcal C_2$ will then produce a state $\ket{\Phi'} = \mathcal{W}_{\mathcal C_2}^{-1}\ket\Phi$ which satisfies
$
\braket{1, \downarrow}{\Phi'} =
\braket{n, \uparrow}{\Phi'} = 0
$,
and is therefore itself the output of a step of quantum walk evolution.

\subsection{3-step reachability conditions}
\label{sec:3step_reachability}
Let us now consider the state $\ket*{\Psi^{(3)}} \equiv \mathcal W_{\mathcal C_3} \ket*{\Psi^{(2)}}$.
The orthogonality relation satisfied by the amplitudes of $\ket*{\Psi^{(2)}}$, that is,
$\bs v_1^{(2)\dagger} \bs v_2^{(2)} = 0$,
can be rewritten as
\begin{equation}
\begin{gathered}
	0 =
	\bs v_1^{(2)\dagger} \bs v_2^{(2)} = 
	\begin{pmatrix}
		u^{(2)*}_{1,\uparrow} & u^{(2)*}_{2,\downarrow}
	\end{pmatrix}
	\begin{pmatrix}
		u^{(2)}_{2,\uparrow} \\ u^{(2)}_{3,\downarrow}
	\end{pmatrix}
	\\
	=
	\begin{pmatrix}
		u^{(2)*}_{1,\uparrow} & u^{(2)*}_{1,\downarrow}
	\end{pmatrix}
	\begin{pmatrix}
		u^{(2)}_{2,\uparrow} \\ u^{(2)}_{2,\downarrow}
	\end{pmatrix}
	+
	\begin{pmatrix}
		u^{(2)*}_{2,\uparrow} & u^{(2)*}_{2,\downarrow}
	\end{pmatrix}
	\begin{pmatrix}
		u^{(2)}_{3,\uparrow} \\ u^{(2)}_{3,\downarrow}
	\end{pmatrix},
\end{gathered}
\end{equation}
where we used
$u^{(2)}_{1,\downarrow} = u^{(2)}_{3,\uparrow} = 0$.
Applying \cref{eq:evolution_equation_with_C} on the above we get
\begin{equation}
\begin{aligned}
	\bs v_1^{(3)\dagger} \mathcal C_3 \mathcal C_3^{-1} \bs v_2^{(3)} +
	\bs v_2^{(3)\dagger} \mathcal C_3 \mathcal C_3^{-1} \bs v_3^{(3)}
	\\ =
	\bs v_1^{(3)\dagger} \bs v_2^{(3)} +
	\bs v_2^{(3)\dagger} \bs v_3^{(3)} = 0.
\end{aligned}
\label{eq:last_condition_for_psi3}
\end{equation}

\begin{table}
\begin{tabular}{cc@{\quad}l}
	\toprule
	\textbf{State} & \textbf{Occupied sites} & \textbf{Constraints} \\
	\midrule \\
	$\ket*{\Psi^{(0)}}$ & 1 & none \\
	\addlinespace[2pt]
	$\ket*{\Psi^{(1)}}$ & 2 & $u_{1,\downarrow} = u_{2,\uparrow} = 0$ \\
	\addlinespace[5pt]
	$\ket*{\Psi^{(2)}}$ & 3 &
	$\begin{dcases}
		u_{1,\downarrow} = u_{3,\uparrow} = 0 \\
		\bs v_1^{\dagger} \bs v_2 = 0
	\end{dcases}$ \\
	\addlinespace[5pt]
	$\ket*{\Psi^{(3)}}$ & 4 &
	$\begin{dcases}
		u_{1,\downarrow} = u_{4,\uparrow} = 0 \\
		\bs v_1^{\dagger} \bs v_3 = 0 \\
		\bs v_1^{\dagger} \bs v_2 + \bs v_2^{\dagger} \bs v_3 = 0
	\end{dcases}$ \\\addlinespace[4pt]
	\vdots & \vdots & \qquad\quad\vdots \\
	$\ket*{\Psi^{(n)}}$ & $n$+1 &
	$\begin{dcases}
		u_{1,\downarrow} = u_{n + 1,\uparrow} = 0 \\
		\sum_{i=1}^s \bs v_i^{\dagger} \bs v_{n-s+i} = 0
	\end{dcases}$ \\
	\bottomrule
\end{tabular}
\caption{
	Summary of the conditions characterizing the states at the various stages of the evolution.
	For better clarity, we have avoided the use of superscripts on the amplitudes.
The amplitudes in each row refer to the corresponding state at that step.
	The conditions shown in the case $\ket*{\Psi^{(n)}}$ hold for all $s=1,..., n-1$, consistently with \cref{eq:general_reachability_conditions}.
}
\label{table:constraints}
\end{table}
Finally, following a reasoning similar to the one used in the previous section, we conclude that the amplitudes of $\ket*{\Psi^{(3)}}$ satisfy three conditions:
the vanishing of the extremal amplitudes, $\bs v_1^{(3)\dagger} \bs v_3^{(3)} = 0$, and~\cref{eq:last_condition_for_psi3}.

\subsection{\texorpdfstring{$n$-step}{n-step} reachability conditions}
The same technique used in the above sections, applied iteratively, leads to the following set of conditions characterizing the amplitudes of $\ket*{\Psi^{(n)}}$:
\begin{equation}
	\sum_{i=1}^s \bs v_i^{(n)\dagger} \bs v^{(n)}_{n-s+i} = 0,
	\text{ for every } s=1,..,n-1,
	\label{eq:general_reachability_conditions}
\end{equation}
plus the vanishing conditions on the extremal elements.
Indeed, if $\ket*{\Psi^{(n)}}$ is the output of $n$ steps, then $\ket*{\Psi^{(n)}} = \mathcal W_{\mathcal C}\ket*{\Psi^{(n-1)}}$ with the amplitudes of $\ket*{\Psi^{(n-1)}}$ satisfying
$\sum_{i=1}^s \bs v_i^{(n-1)\dagger} \bs v_{(n-1)-s+i}^{(n-1)} = 0$
for all $s=1,...,n-2$,
and $u_{1,\downarrow}^{(n-1)} = u_{n, \uparrow}^{(n-1)} = 0$.
These last two conditions give, following the same reasoning used in~\cref{sec:2step_reachability}, the orthogonality condition $\bs v_1^{(n)\dagger} \bs v_n^{(n)} = 0$.
On the other hand, the same reasoning used in~\cref{sec:3step_reachability} can be used to derive the $(s+1)$-th equation in~\cref{eq:general_reachability_conditions} from the $s$-th one on $\ket*{\Psi^{(n-1)}}$.

While in the above we considered only the case of a quantum walk in which the initial walker state was $\ket1$, 
the results hold in the more general case of a generic initial walker state, possibly spanning more than one position.
Indeed, \cref{eq:general_reachability_conditions} is equivalent to stating that $\ket*{\Psi^{(n)}}$ is the output of (at least) $n$ steps of quantum walk evolution, \textit{regardless of the initial state}.
We refer to~\cref{app:reachability_conditions} for a more detailed discussion about this more general case.

In \mbox{\cref{table:constraints}} we give a summary of the conditions characterizing the states at different steps.
The total number of (real) degrees of freedom of a walker state after $n$ steps, considering the two vanishing extremal amplitudes, is $4n - 2$.
Additionally, the set of constraints in \cref{eq:general_reachability_conditions} amounts to $2(n-1)$ real conditions.
It follows that the space of reachable states $\ket*{\Psi^{(n)}}$ has dimension $2n$, to compare with the number of degrees of freedom of a general state living in the same Hilbert space, that is, $4n + 2$.

\subsection{Set of coin operators generating a state}
\label{sec:coin_operators_generating_state}
We conclude our analysis of the global reachability conditions by remarking that \cref{eq:general_reachability_conditions} completely characterises the set of quantum states reachable after $n$ steps.
In other words, not only every state that is the result of a quantum walk evolution satisfies it, but also for every state $\ket*{\Phi}$ that satisfies \cref{eq:general_reachability_conditions}, there is a set of coin operators $\{ \mathcal C_i \}_{i=1}^n$ and an initial state $\ket{\Phi_{in}}$ such that
$\ket\Phi = 
	\mathcal W_{\mathcal C_n} \cdots \mathcal W_{\mathcal C_1}
	\ket{\Phi_{in}}$.
In fact, given the state $\ket\Phi$, this set of coin operators is efficiently computable.
This was already sketched in a previous section, but we will here give a more detailed and general description of the computation.
We start by noting that,
denoting with $\bs v_i$ the $\bs v$-vectors built from the amplitudes of a generic $\ket{\Phi}$ satisfying~\cref{eq:general_reachability_conditions}, we have $\bs v_1^\dagger \bs v_n = 0$.
This immediately implies the existence of a \dimenst{2}{2} unitary matrix $\mathcal C_n$, and complex numbers $a$ and $b$, such that
\begin{equation}
	\bs v_1 = \mathcal C_n
	\begin{pmatrix} a \\ 0 \end{pmatrix},
	\qquad
	\bs v_n = \mathcal C_n
	\begin{pmatrix} 0 \\ b \end{pmatrix}.
	\label{eq:equations_giving_C}
\end{equation}
The first of the above equations implies that the second row of $\mathcal C_n^{-1}$ is orthogonal to $\bs v_1$, that is,
$
	- c_{21} u_{1, \uparrow} + c_{11} u_{2, \downarrow} = 0.
$
This condition is enough, together with the constraint of the columns of a unitary matrix being normalised, to determine the first column of $\mathcal C_n$ up to a phase.
The orthonormality constraint that $\mathcal C_n$ must satisfy then determines the elements of the second column, again up to a phase.
This leaves the freedom to choose two different phases for the two columns.
Finally, remembering that the global phase of $\mathcal C_n$ does not have physical consequences, we conclude that $\mathcal C_n$ is determined up to a phase difference between the two columns.
More explicitly, the above argument tells us that the coin operators generating a walker state spanning $n+1$ sites are all and only the unitaries of the form
\begin{equation*}
\begin{aligned}
	\mathcal C_n
	&= N \begin{pmatrix}
		u_{1,\uparrow} & e^{i\alpha} u_{\makebox[0pt][l]{$\scriptstyle n$}\phantom{n+1},\uparrow} \\
		u_{2,\downarrow} & e^{i\alpha} u_{n+1,\downarrow}
	\end{pmatrix} = N \begin{pmatrix}
		u_{1,\uparrow} & -e^{i\alpha} u_{2,\downarrow}^* \\
		u_{2,\downarrow} & \phantom{-}e^{i\alpha} u_{1,\uparrow}^*
	\end{pmatrix},
\end{aligned}
\end{equation*}
with $\alpha$ arbitrary and $N$ normalization constant.
This coin operator can then be written, using for example the parametrization of~\cite{chandrashekar2008optimizing}, as
\begin{equation}
	\mathcal C_n = 
	\begin{pmatrix}
		e^{i\xi} \cos\theta &
		e^{i\zeta} \sin\theta \\
		-e^{-i\zeta}\sin\theta & e^{-i\xi}\cos\theta
	\end{pmatrix},
	\label{eq:su2_coin_matrix}
\end{equation}
with $\theta\in [0, \pi/2]$, $\xi$ and $\zeta$ satisfying
$\tan\theta = \lvert u_{2,\downarrow} \rvert/\lvert u_{1,\uparrow} \rvert$
and
$\xi + \zeta \pm \pi = \arg\left(u_{1, \uparrow}/u_{2, \downarrow}\right)$.
The freedom in $\alpha$ is here translated into the phase difference $\zeta - \xi$ not being uniquely determined.
More explicitly, when tracing back the evolution of the walker one can at every step write the coin operator as in \cref{eq:su2_coin_matrix} with $\xi \to \xi + \varphi$ and $\zeta \to \zeta - \varphi$ for an arbitrary $\varphi\in \mathbb R$.
Trivial modifications must be applied to the above formulae when $u_{2,\downarrow} = 0$.

To obtain the coin operators at the previous steps, we use the $\mathcal C_n$ computed above to get the amplitudes of $\mathcal W_{\mathcal C_n}^{-1} \ket\Phi$.
These amplitudes will again satisfy the orthogonality condition between the $\bs v$-vectors at the first and last sites:
${\bs v_1^\dagger \bs v_{n-1} = 0}$
(where now $\bs v_i$ refers to the amplitudes after the $(n-1)$-th step).
The same argument used above can now be repeated, to compute the coin operator at this step.
Iterating this procedure we can compute all of the coin operators, with the exception of the first one.
In fact, the amplitudes after the first step do not satisfy the orthogonality condition like \cref{eq:equations_giving_C} anymore (the only condition on them is the vanishing of the extremal amplitudes), so that the above argument breaks.
This last (first) coin operator is nevertheless easily solved for by remembering that the only non vanishing amplitudes at this stage of the evolution are $u_{1,\uparrow}$ and $u_{2, \downarrow}$, so that if the initial state is
${\ket{\Phi_{in}} = u^{(0)}_{1,\uparrow} \ket{1, \uparrow} + u^{(0)}_{1, \downarrow} \ket{1, \downarrow}}$, then the first coin operator must satisfy
\begin{equation}
	\begin{pmatrix}
		u_{1, \uparrow} \\ u_{2,\downarrow}
	\end{pmatrix}
	= \mathcal C_1
	\begin{pmatrix}
		u^{(0)}_{1, \uparrow} \\ u^{(0)}_{1, \downarrow}
	\end{pmatrix}.
\end{equation}
The values of $u_{1, \uparrow}, u_{2,\downarrow}, u^{(0)}_{1, \uparrow}$ and $u^{(0)}_{1, \downarrow}$ are known, so that this equation is readily solved for the elements of $\mathcal C_1$, for any initial state $\ket{\Phi_{in}}$.

\section{Focusing on walker states}
\label{sec:focusing_walker_states}
\Cref{sec:reachability} focused on characterising the set of quantum states that can be generated by a quantum walk evolution.
We here instead consider a different scenario, in which one is interested in generating a target \emph{qudit} state over the walker's degree of freedom, as opposite to wanting to generate a target quantum state in the full walker+coin space.
In other words, we fix a number of steps $n$ and a target superposition over the sites $\ket\phi=\sum_{i=1}^{n+1} u_i\ket i$, and ask whether there is a combination of coin operators $\{\mathcal C_i\}_{i=1}^{n}$, a coin state $\ket\gamma$, and an initial state $\ket{\Psi_0}=\ket{1, \alpha}$ for some initial coin state $\ket\alpha$, such that (up to a normalisation factor)
$\mel{\gamma}{\mathcal W_{\mathcal C_{n}} \cdots\mathcal W_{\mathcal C_{1}}}{\Psi_0} = \ket\phi$.
In particular, we are interested in finding whether an arbitrarily chosen superposition of sites $\ket\phi$ can be generated in this way.
The main tool that will be used to answer this question is the set of conditions developed in~\cref{sec:reachability}.
We will find the answer to be always positive, provided special degenerate conditions are satisfied.
For the analytical results we will focus on the case of the coin being projected over $\ket+$,
and provide numerical results for the more general case.

Let $\ket\phi$ be an arbitrary superposition of $n+1$ walker sites with amplitudes $u_k \equiv \braket{k}{\phi}$.
We want to find a reachable walker state $\ket\Phi$, in the full walker+coin space, such that
$\ket{+}_c \!\braket{+}{\Phi} \propto \ket\phi \!\otimes\! \ket+$.
Denoting with $u_{i,s}$ the amplitudes of $\ket\Phi$, this amounts to finding a set of amplitudes $\{u_{i,s}\}$ that simultaneously satisfies
${N(u_{i,\uparrow}+u_{i,\downarrow})= u_i}$
and the reachability conditions of~\cref{eq:general_reachability_conditions}.
To do this, we parametrize the set of all walker states $\ket\Phi$ whose amplitudes give the target after projection as
\begin{equation}
\begin{aligned}
	\ket\Phi &= N \Big( u_1 \ket{1,\uparrow} + u_{n+1} \ket{n+1,\downarrow} \\
	&+ \sum_{i=2}^n \big[
		(u_i - d_i) \ket{i, \uparrow} +
		d_i \ket{i, \downarrow}
	\big] \Big),
\end{aligned}
\label{eq:parametrized_state_with_ds}
\end{equation}
where $\{ d_i \}_{i=2}^n$ is a set of parameters to be determined, and $N$ is a normalization constant.
Projecting \cref{eq:parametrized_state_with_ds} over the coin state $\ket+$ gives the correct result for all values of $d_i$.
The problem is therefore reduced to that of finding a set $\{d_i\}_i$ corresponding to a reachable state as per \cref{eq:general_reachability_conditions}.
Direct substitution of \cref{eq:parametrized_state_with_ds} into \cref{eq:general_reachability_conditions} gives the set of conditions that these parameters have to satisfy
\begin{equation}
	\sum_{i=1}^s
	(u_i - d_i)^*
	(u_{n-s+i} - d_{n-s+i})
	+
	d_{i+1}^* d_{n-s+i+1}
	= 0,
	\label{eq:conditions_for_ds}
\end{equation}
for every $s=1,...,n-1$,
with $d_1=0$ and $d_{n+1}=u_{n+1}$.
Splitting real and imaginary parts, \cref{eq:conditions_for_ds} is equivalent to a system of $2(n-1)$ real quadratic equations in $2(n-1)$ real variables.
It follows that~\cref{eq:conditions_for_ds} has solutions for almost all target states, except for a subset of states of measure zero.
In the simplest case $n=2$ an explicit solution can be written as
\begin{equation}
	d_2 = \frac{
		u_1(u_1^* u_2 + u_2^* u_3)
	}{
		\lvert u_1 \rvert^2 - \lvert u_3 \rvert^2
	}~\text{ for}~\lvert u_1 \rvert \neq \lvert u_3 \rvert.
	\label{eq:solution_for_d2}
\end{equation}
Note that this does not imply that for $\lvert u_1 \rvert = \lvert u_3 \rvert$ there is no solution for $d_2$, but only that \cref{eq:solution_for_d2} does not apply in that case.
More generally, \cref{eq:conditions_for_ds}
can be solved numerically with ease for small $n$, giving multiple solutions for any randomly selected target state.
As remarked above for the general case, it is still possible for a solution of~\cref{eq:conditions_for_ds} to not exist, provided some specific degenerate conditions are met.
Further analysis of the analytical solution for two steps is provided in~\cref{app:analytical_sol_2steps}.

A solution of~\cref{eq:conditions_for_ds}, once found, can be used to compute the coin parameters producing a target $\ket\Phi$ as shown in~\cref{sec:coin_operators_generating_state}.
This effectively allows to generate superpositions of $n+1$ sites using $n$ steps of quantum walk evolution, via projection of the coin state over $\ket+$ at the end of walk.
However, the projection makes this scheme probabilistic, so it is important to ensure that the generation probabilities are not vanishingly small.
We tested this with up to 5 steps analysing the solutions of \cref{eq:conditions_for_ds}, and with up to 20 steps finding the coin parameters generating target states using a more general numerical maximization algorithm.
The results of these analyses are given in~\cref{sec:numerical_solution_reachability_conditions,sec:numerical_fid_max},
where we also consider the approximate (namely, with fidelity smaller than 1) engineering of target states,
and the robustness of the method with respect to imperfections of the coin parameters.
We will also find that in the vast majority of instances the numerical maximization identifies coin parameters able to generate arbitrary target states with both high probabilities and fidelities.
\footnote{The code used to produce these results is freely available at \href{https://github.com/lucainnocenti/QSE-with-QW-code}{https://github.com/lucainnocenti/QSE-with-QW-code}.}

\subsection{Numerical solution of reachability conditions}
\label{sec:numerical_solution_reachability_conditions}
\begin{figure*}[]
	\centering
	\begin{minipage}[b]{0.5\textwidth}
		\begin{tikzpicture}
			\node (img) {\includegraphics[width=\columnwidth]%
				{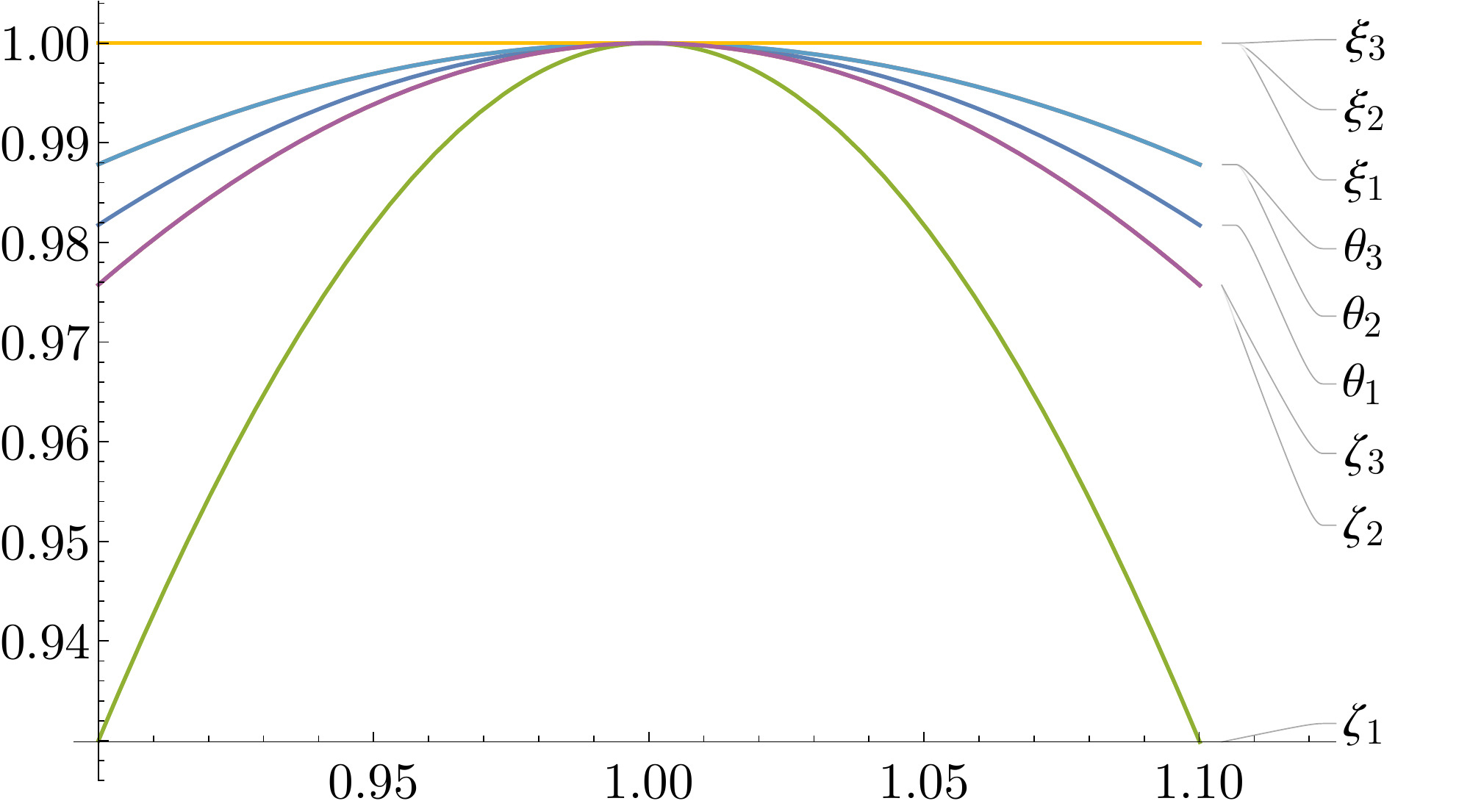}
			};
			\node [overlay] (F) at (-3.8, 2.8) {\scalebox{1.2}{$\mathcal F$}};
		\end{tikzpicture}
	\end{minipage}%
	\begin{minipage}[b]{0.5\textwidth}
		\begin{tikzpicture}
			\node (img) {\includegraphics[width=\columnwidth]%
				{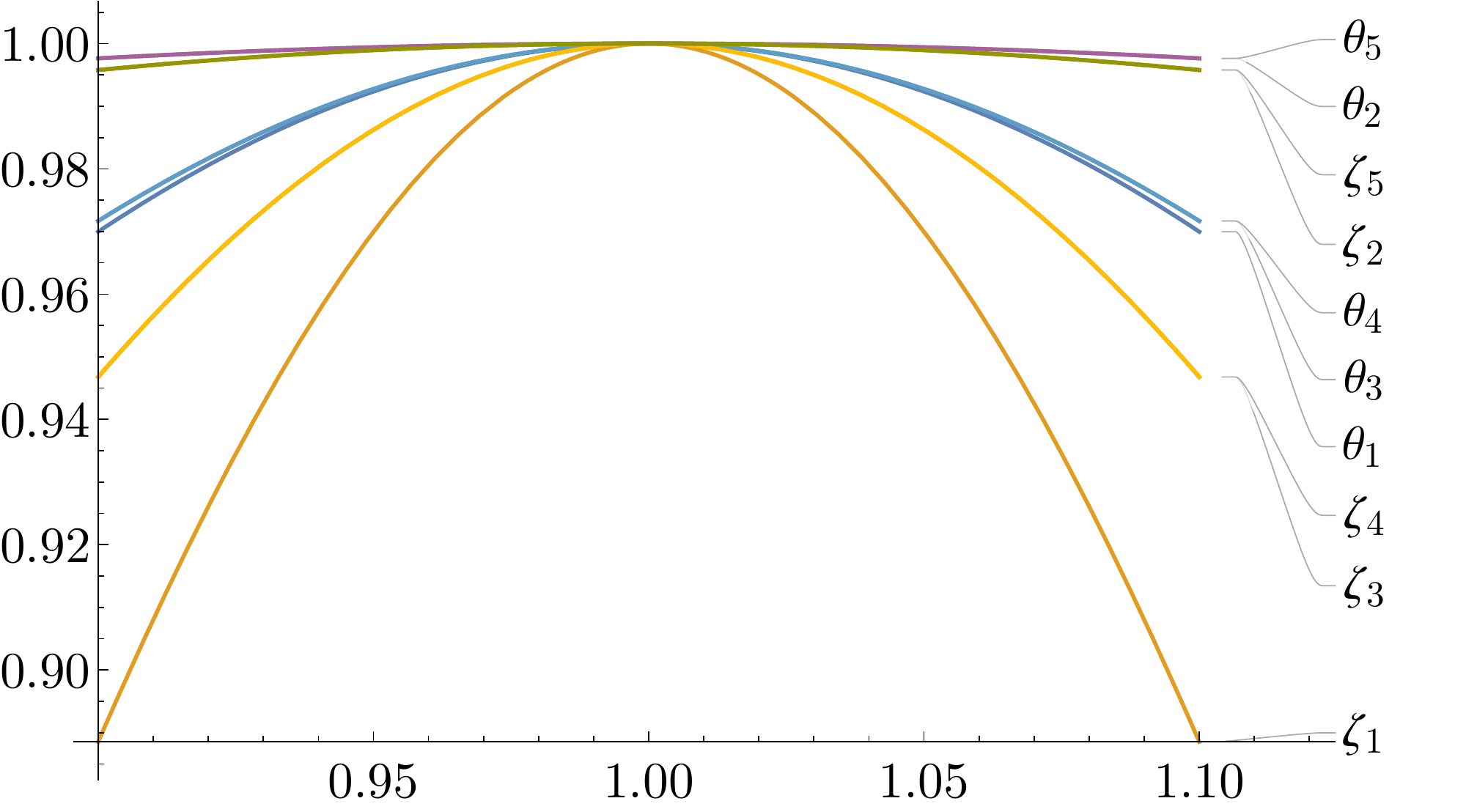}
			};
			\node [overlay] (F) at (-3.8, 2.8) {\scalebox{1.2}{$\mathcal F$}};
		\end{tikzpicture}
	\end{minipage}
	\caption{
		Fidelity varying the various coin parameters for 3 (left) and 5 (right) steps, when the target is the completely balanced superposition over 4 and 6 modes, respectively.
		The variation of the coin parameters is here shown in percentage: the edges of the plots correspond to a variation of 10\% of a single parameter with the others kept fixed at their optimal value.
	}
	\label{fig:stabilities_3and5steps_balanced_target}
\end{figure*}

We solved numerically~\cref{eq:conditions_for_ds} for $n=2,3,4$ and $5$ steps, for a number of random target states sampled from the uniform Haar distribution.
These equations resulted in a varying number of solutions for different target states and number of steps:
always $1$ solution for 2 steps, $2$ or $4$ solutions for 3 steps, $3, 5, 7$ or $9$ solutions for 4 steps,
and $6, 8, 10, 12, 14$ or $16$ solutions for 5 steps.
Different solutions for the same target state correspond to different dynamics before the projection and different projection probabilities.
This point is illustrated in \cref{fig:minmax_probabilities}, where we reported maximum and minimum projection probabilities for each randomly generated target state.
From these results the advantage of solving~\cref{eq:conditions_for_ds} is clear: having the whole set of solutions we can choose the more convenient one in terms of projection probability.
Having access to the various solutions generating a given target state, we can also study their stability properties.
As shown in \cref{fig:stabilities_5steps}, a general trend is that sets of coin parameters associated to smaller projection probabilities are also less stable, in the sense that a small perturbation of a coin parameter leads to a state significantly different than the target one.
As an example, we can use this method to generate a balanced superposition over 4 sites (using therefore 3 steps): $\ket\phi = (1, 1, 1, 1) / 2$.
This results in two solutions for the $d$: $\bs d = (-i, 1+i)/2$ and $\bs d = (i, 1 - i)/2$,
corresponding to the full states
\begin{equation*}
\begin{split}
	\frac{1}{2}\Big(
		&\ket{1,\uparrow} + 
		(1 \mp i) \ket{2,\uparrow} \pm i \ket{2, \downarrow} \\
		&\quad \pm i \ket{3, \uparrow} + (1 \mp i) \ket{3, \downarrow} +
		\ket{4, \downarrow}
	\Big),
\end{split}
\end{equation*}
which both result in a projection probability over $\ket+$ of 1/4.
The same procedure applied to a balanced superposition over 6 sites (5 steps) results in 6 solutions, 2 of which with real $\bs d$ and projection probability $\simeq 0.145$, and the other 4 with complex $\bs d$ and projections probabilities of 1/6.
It is worth noting that while the projection probabilities over balanced superpositions over $n$ sites seem to vanish with $1/2n$, a simple phase change of an element can radically change this probability.
As an example, again the case of 6 sites, if the target state is instead $\ket\phi \simeq (1,1,1,1,1,-1)$ the maximum projection probability becomes $\simeq 0.35$.
The stability of the above solutions for balanced states when a small perturbation is applied to the coin parameters is shown in \cref{fig:stabilities_3and5steps_balanced_target}.

\subsection{Numerical maximization of  fidelity}
\label{sec:numerical_fid_max}
The numerical solution of~\cref{eq:conditions_for_ds} for more than 5 steps is computationally difficult, due to the complexity of the resulting system of equations,
and we therefore employed a different numerical technique for this regime.
We wrote the fidelity for a given target state as a function of the coin parameters, and found the set of parameters maximizing such fidelity using a standard optimization algorithm.
In this way we could probe instances with up to 20 steps much more efficiently than we could have done by solving~\cref{eq:conditions_for_ds}.
Furthermore, this method eases the study of different final projections, and allows to include the parameters of the projection itself in the optimization.

\begin{figure*}[]
	\centering
	\begin{minipage}[b]{0.5\textwidth}
		\begin{tikzpicture}
			\node (img) {\includegraphics[width=\columnwidth]%
				{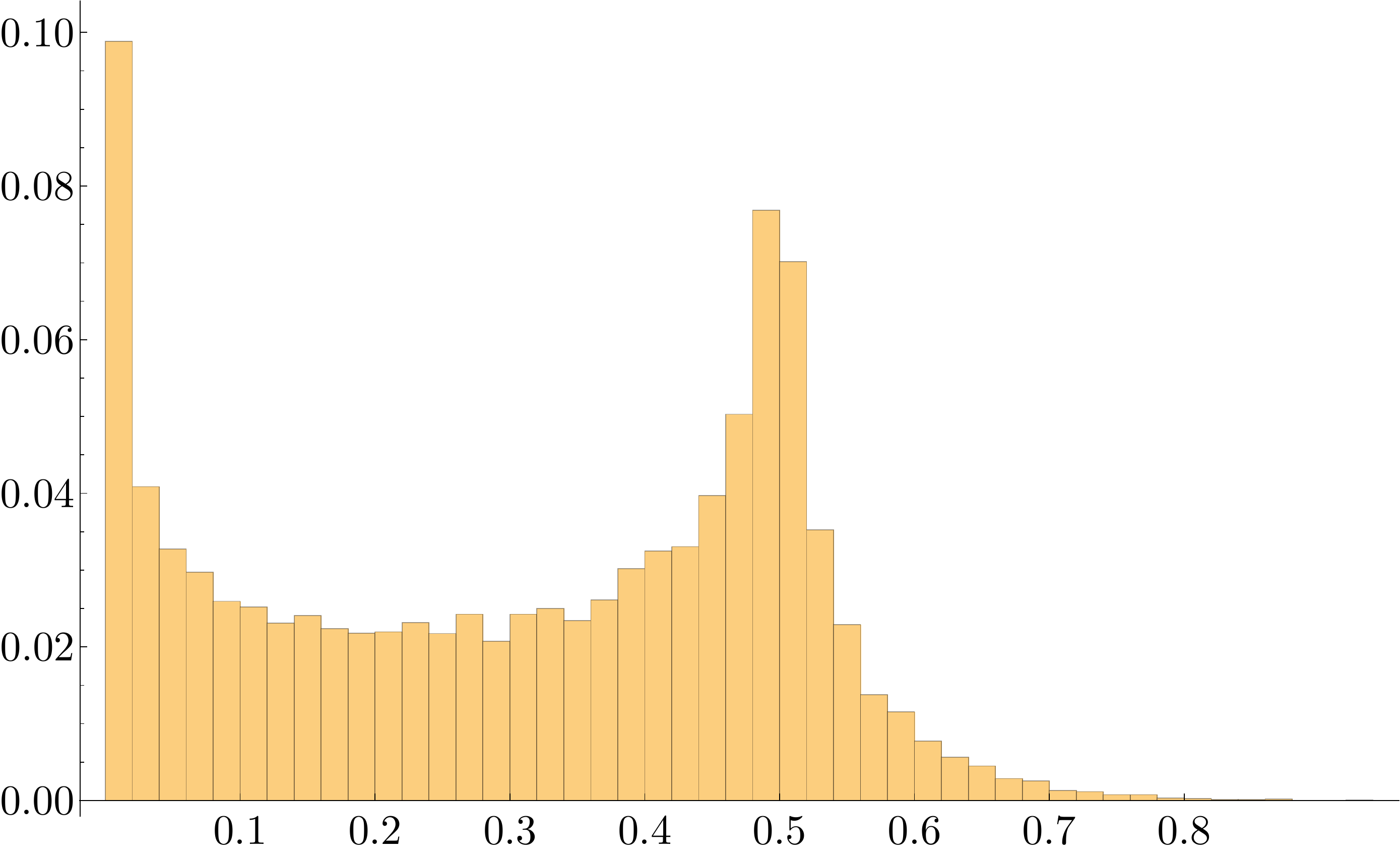}
			};
			\node [overlay] (x-axis) at (3.6, -2.6) {\scalebox{1.2}{$p$}};
			\node [overlay] at (0, 2) {\fbox{2 steps}};
		\end{tikzpicture}
	\end{minipage}%
	\begin{minipage}[b]{0.5\textwidth}
		\begin{tikzpicture}
			\node (img) {\includegraphics[width=\columnwidth]%
				{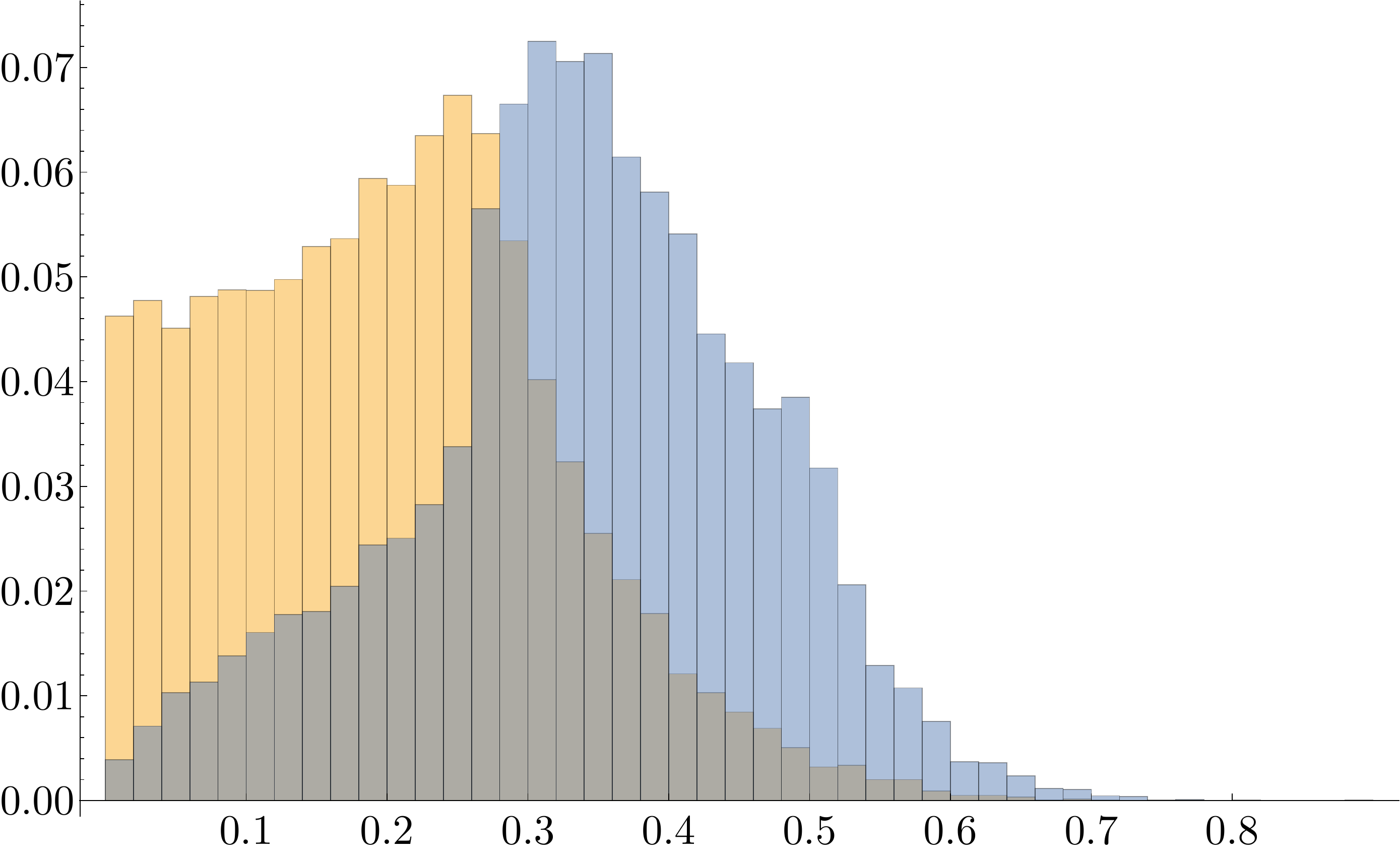}
			};
			\node [overlay] (x-axis) at (4, -2.6) {\scalebox{1.2}{$p$}};
			\node [overlay] at (.2, 2) {\fbox{3 steps}};
		\end{tikzpicture}
	\end{minipage}
	\begin{minipage}[b]{0.5\textwidth}
		\begin{tikzpicture}
			\node (img) {\includegraphics[width=\columnwidth]%
				{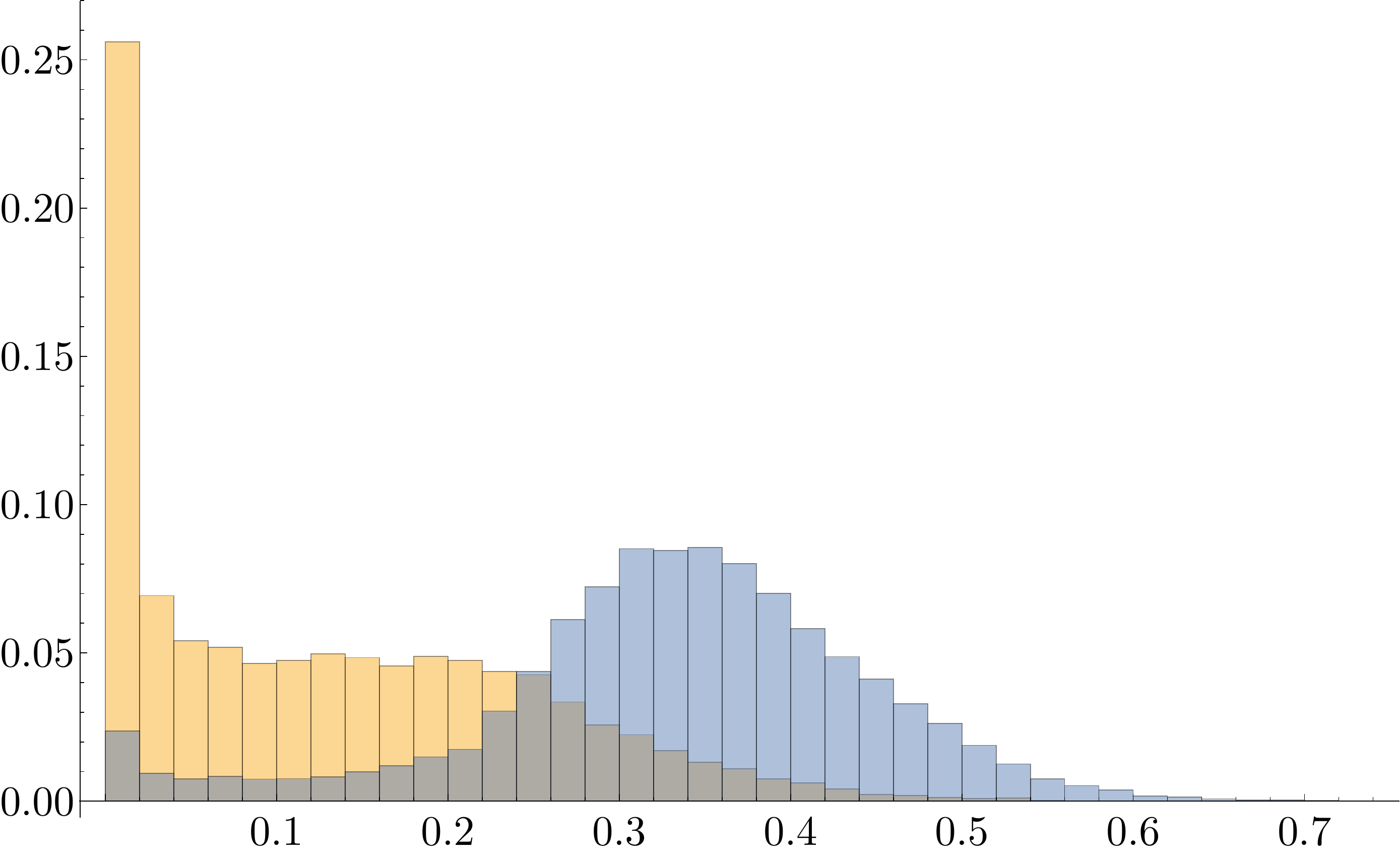}
			};
			\node [overlay] (x-axis) at (4.2, -2.6) {\scalebox{1.2}{$p$}};
			\node [overlay] at (0, 2) {\fbox{4 steps}};
		\end{tikzpicture}
	\end{minipage}%
	\begin{minipage}[b]{0.5\textwidth}
		\begin{tikzpicture}
			\node (img) {\includegraphics[width=\columnwidth]%
				{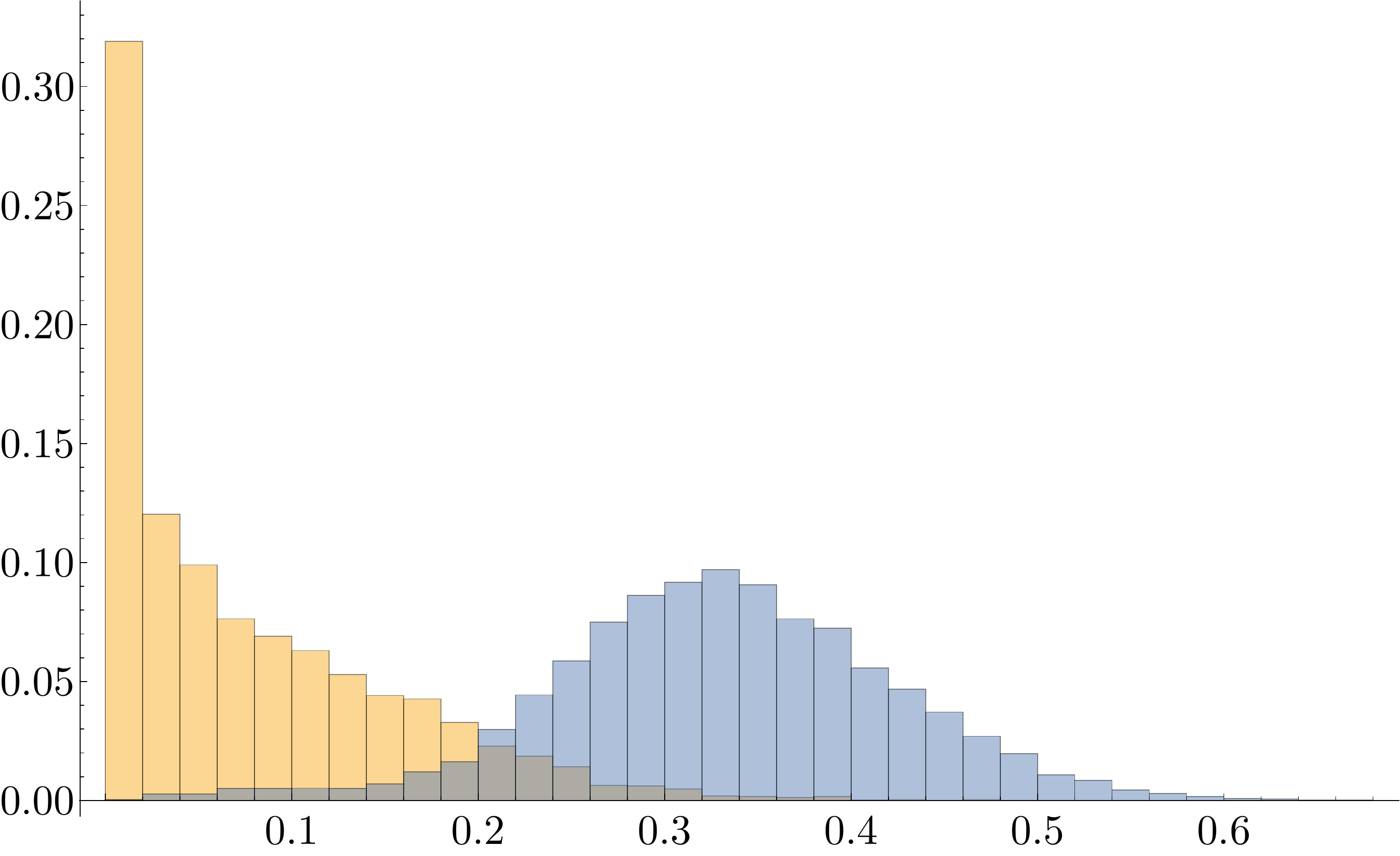}
			};
			\node [overlay] (x-axis) at (4, -2.6	) {\scalebox{1.2}{$p$}};
			\node [overlay] at (0, 2) {\fbox{5\,\, steps}};
		\end{tikzpicture}
	\end{minipage}
	\caption{
		Distribution of projection probabilities computed 1) solving~\cref{eq:conditions_for_ds} for the parameters $d_i$,
		2) computing the projection probabilities associated to each full state given by one such solution set, and
		3) picking the solution set for the $\{d_i\}$ associated to lowest and highest projection probabilities.
		The shown data is for the cases of 2, 3, 4 and 5 steps.
		For 2, 3 and 4 steps the data shows the distribution of probabilities from a sample set of 20000 target states, drawn according to the uniform Haar measure over the set of target states.
		For 5 steps only 6000 states were used (being this case much more computationally expensive).
		On the $y$-axis is shown the fraction of target states in a given bin.
	}
	\label{fig:minmax_probabilities}
\end{figure*}
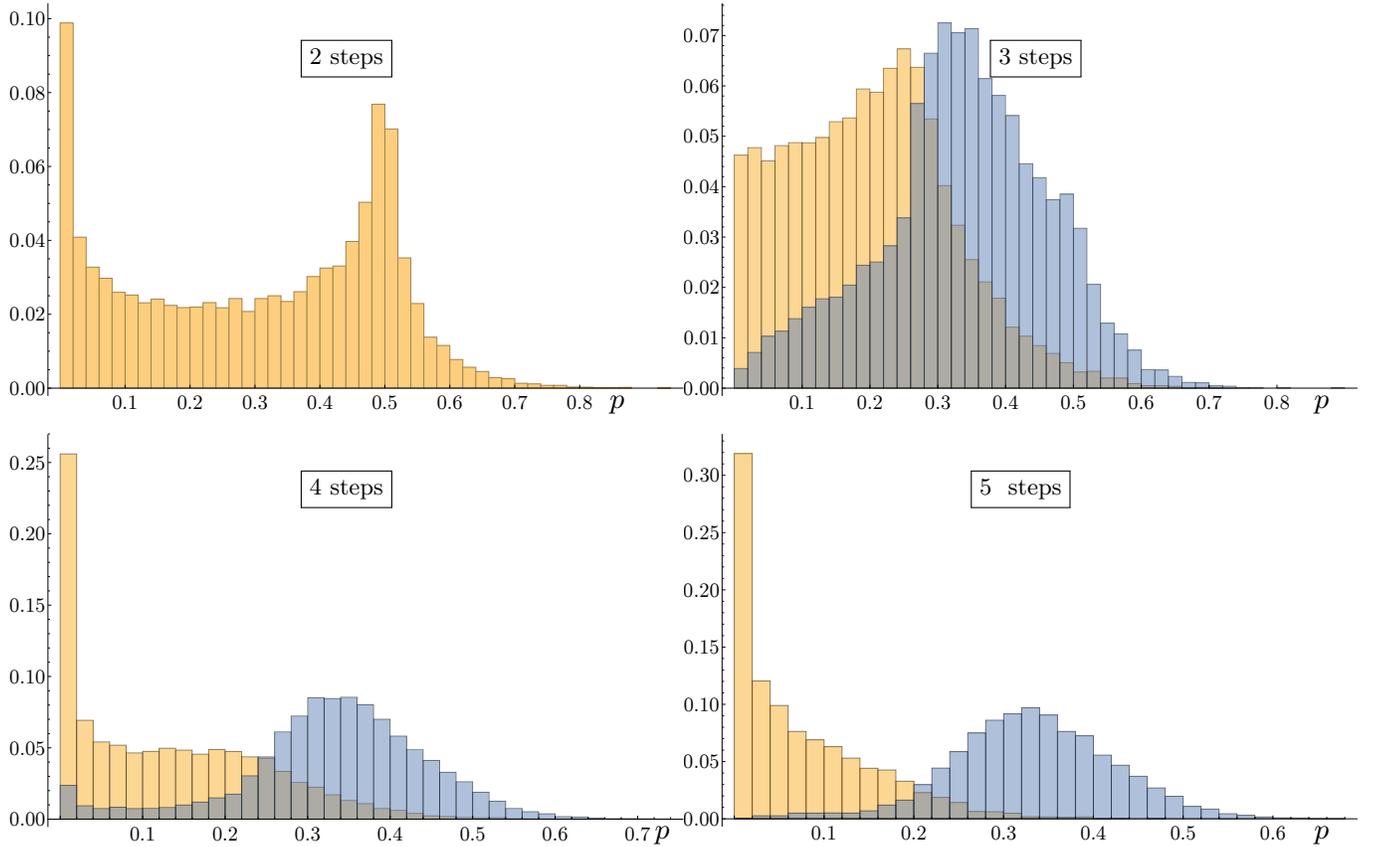

\begin{figure*}[tbp]
	\centering
	\begin{minipage}[b]{0.5\textwidth}
		\includegraphics[width=\columnwidth]
			{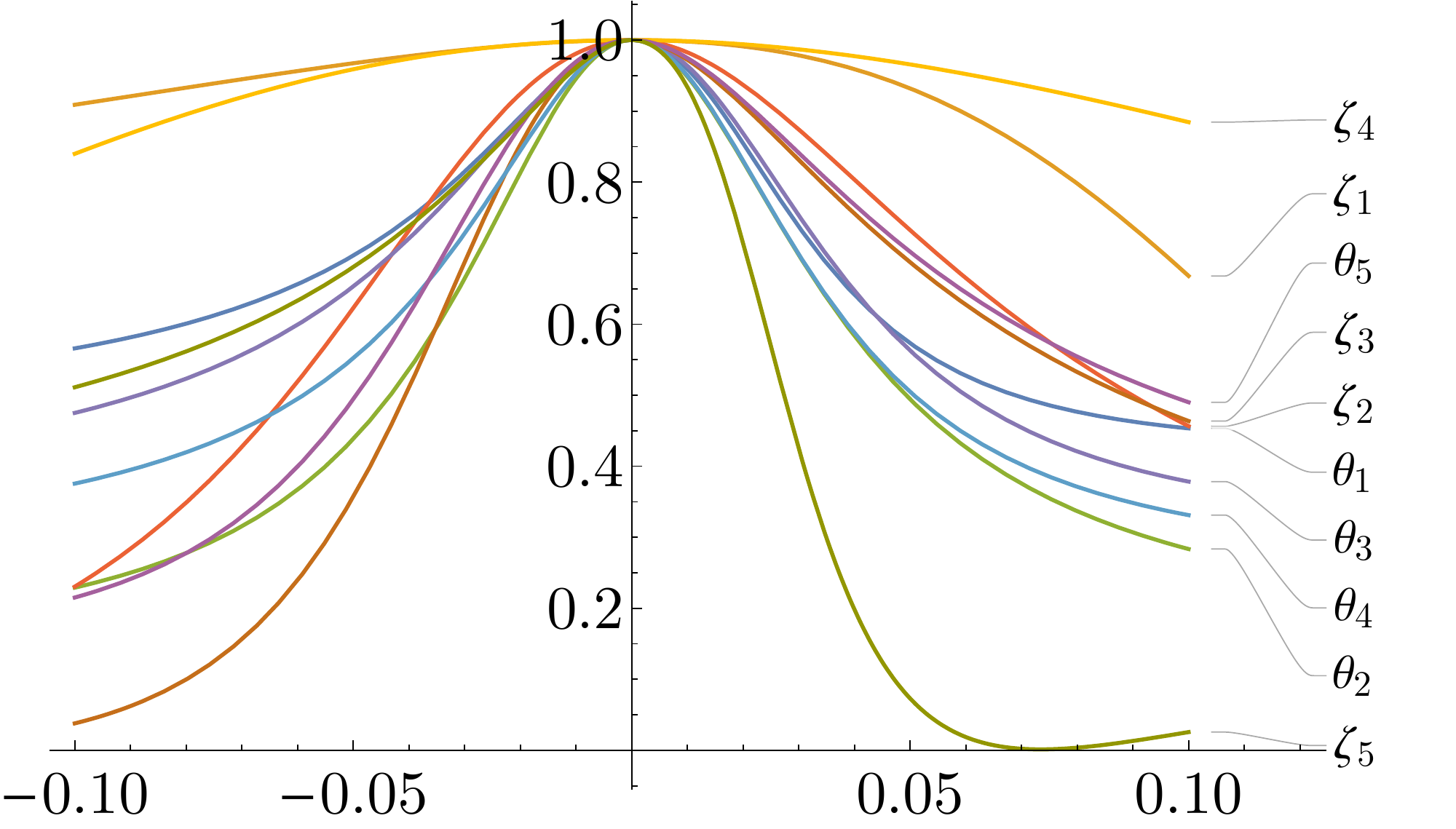}
	\end{minipage}%
	\begin{minipage}[b]{0.5\textwidth}
		\includegraphics[width=\columnwidth]
			{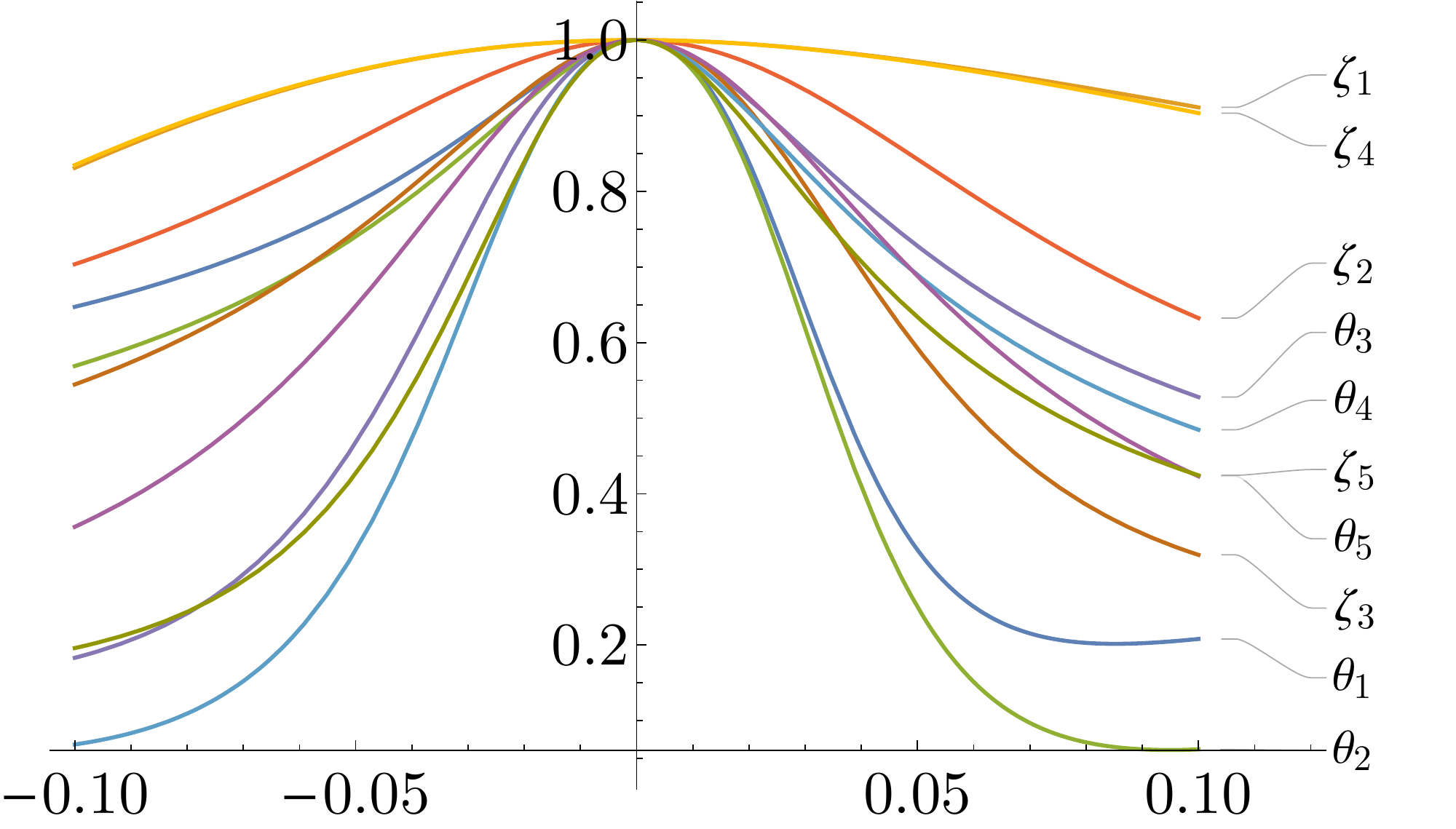}
	\end{minipage}
	\begin{minipage}[b]{0.5\textwidth}
		\includegraphics[width=\columnwidth]
			{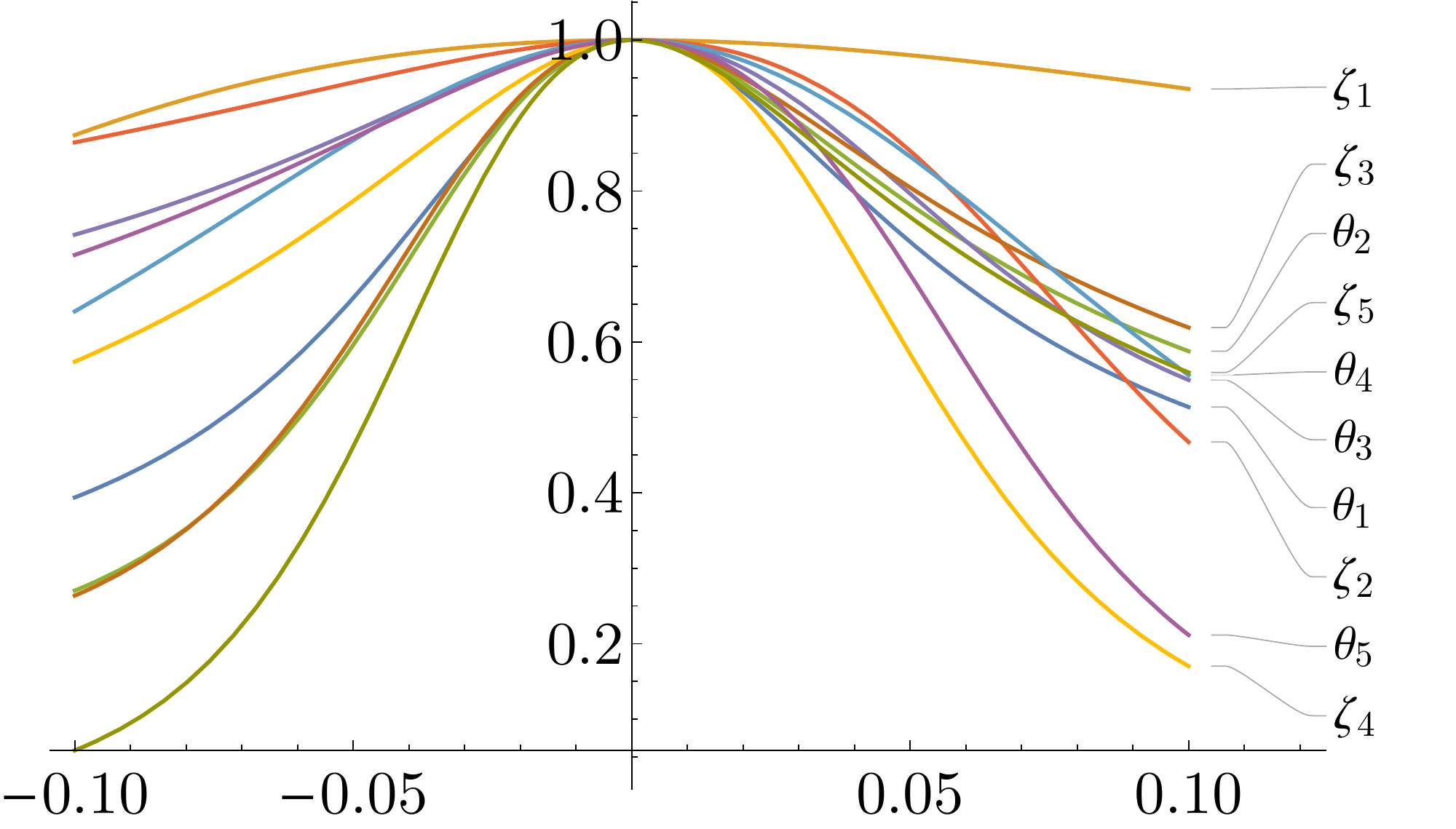}
	\end{minipage}%
	\begin{minipage}[b]{0.5\textwidth}
		\includegraphics[width=\columnwidth]
			{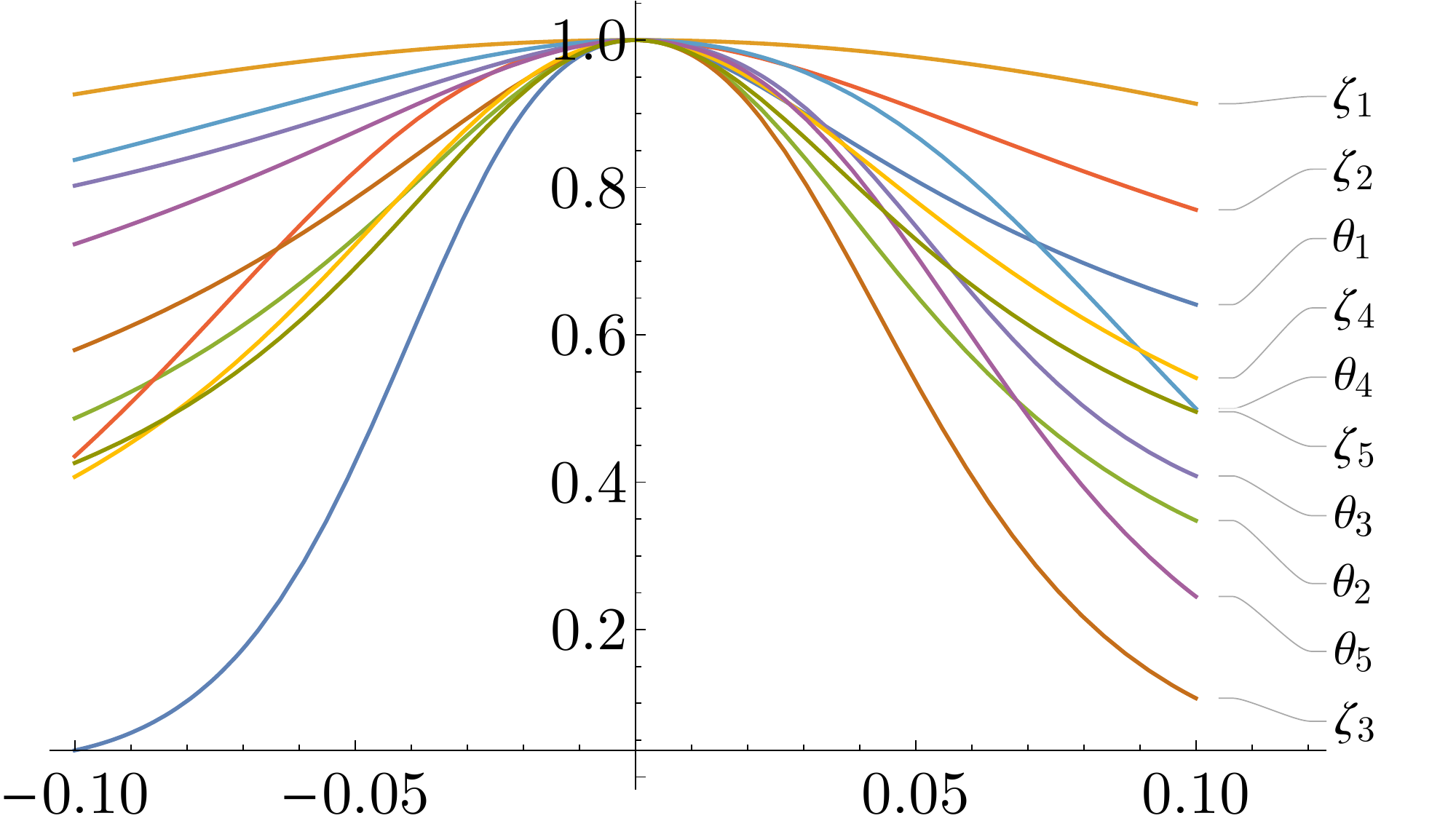}
	\end{minipage}
	\begin{minipage}[b]{0.5\textwidth}
		\includegraphics[width=\columnwidth]
			{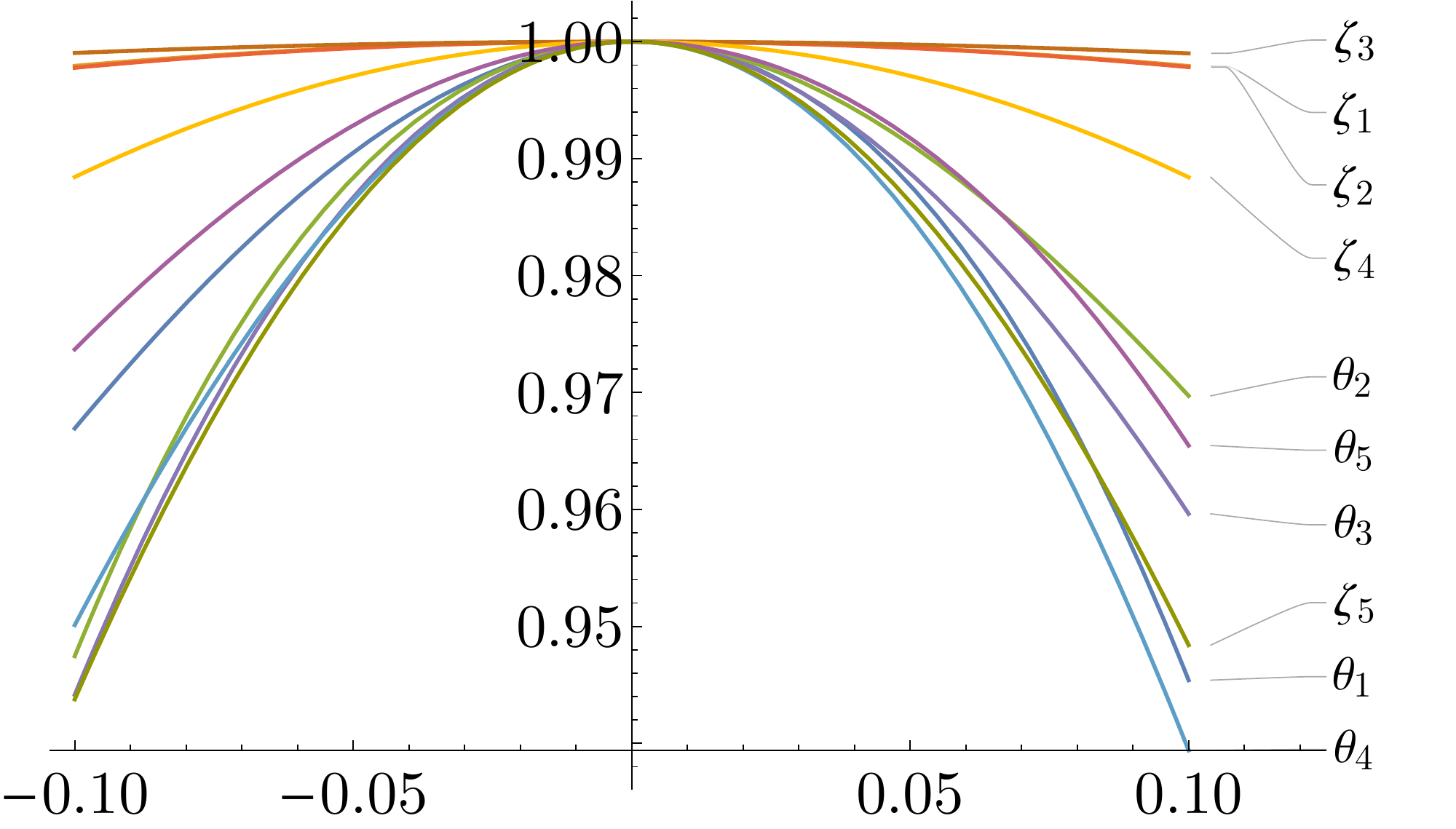}
	\end{minipage}%
	\begin{minipage}[b]{0.5\textwidth}
		\includegraphics[width=\columnwidth]
			{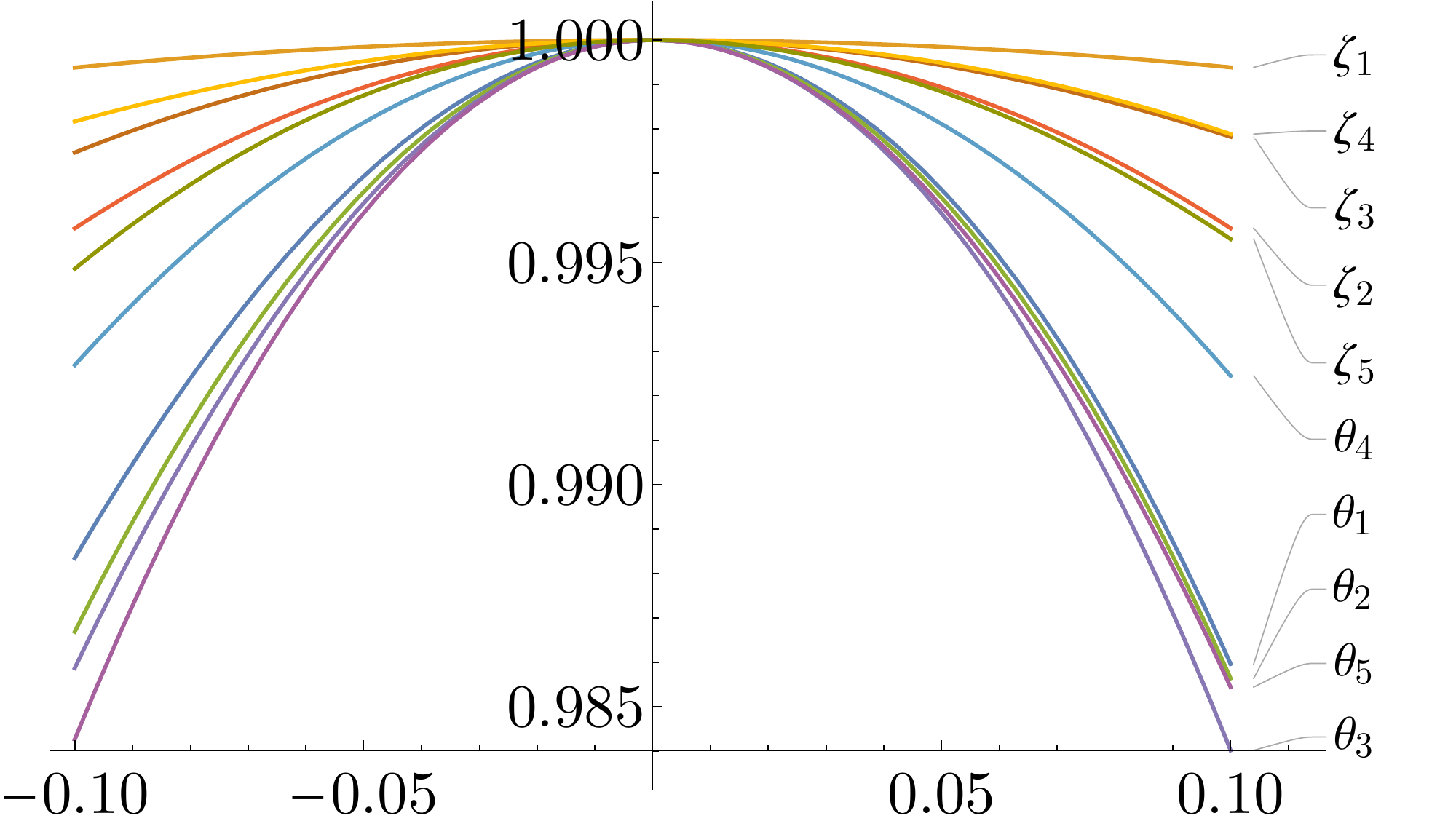}
	\end{minipage}
	\caption{
		Behaviour of projection probability of the solutions found solving~\cref{eq:conditions_for_ds} for the target state $\ket\phi \simeq (0.053, -0.078 + 0.603i, -0.524 + 0.189i, -0.302 + 0.363i, 0.182 + 0.099i, 0.042 - 0.224i)$.
		In each of the plots, the final fidelity is plotted against many coin parameters, each time fixing the value of all of them except for one, whose value is changed by an absolute value $\epsilon$ (x-axis).
		The 6 figures correspond to the 6 solutions for this target state, having respectively the projection probabilities:
		$0.0014$ (top left), $0.0020$ (top right),
		$0.0036$ (middle left), $0.0038$ (middle right),
		$0.14$ (bottom left) and $0.398$ (bottom right).
		As clearly illustrated in this case, solutions with low projection probabilities tend to present an higher degree of instability with respect to small changes of the coin parameters.
	}
	\label{fig:stabilities_5steps}
\end{figure*}

\begin{figure*}[]
	\centering
	\begin{minipage}[b]{0.5\textwidth}
		\begin{tikzpicture}
			\node (img) {\includegraphics[width=\columnwidth]%
				{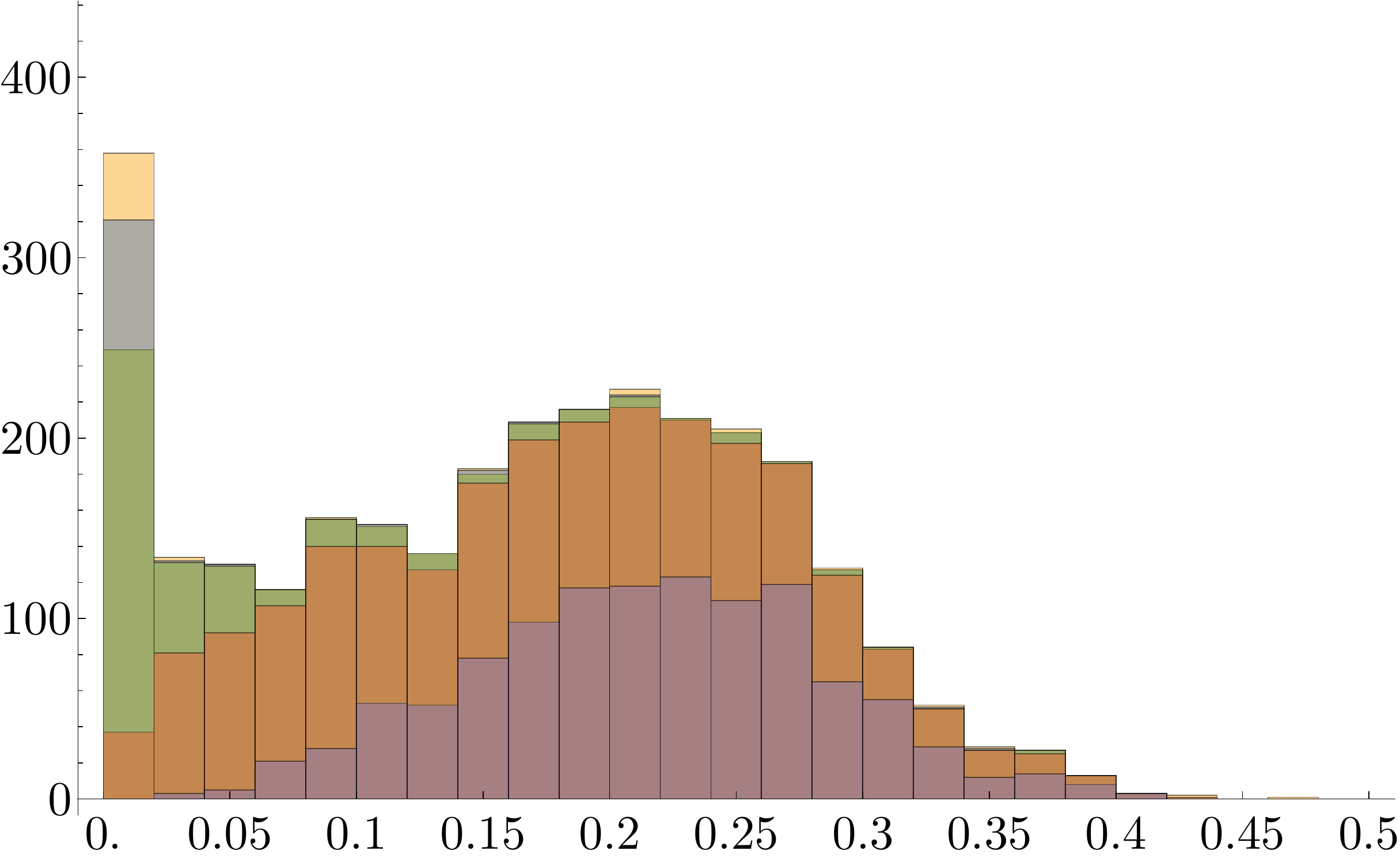}
			};
			\node [overlay] (x-axis) at (3.8, -2.8) {\scalebox{1.2}{$p$}};
			\node [overlay] at (0, 2) {\fbox{15 steps}};
		\end{tikzpicture}
	\end{minipage}%
	\begin{minipage}[b]{0.5\textwidth}
		\begin{tikzpicture}
			\node (img) {\includegraphics[width=\columnwidth]%
				{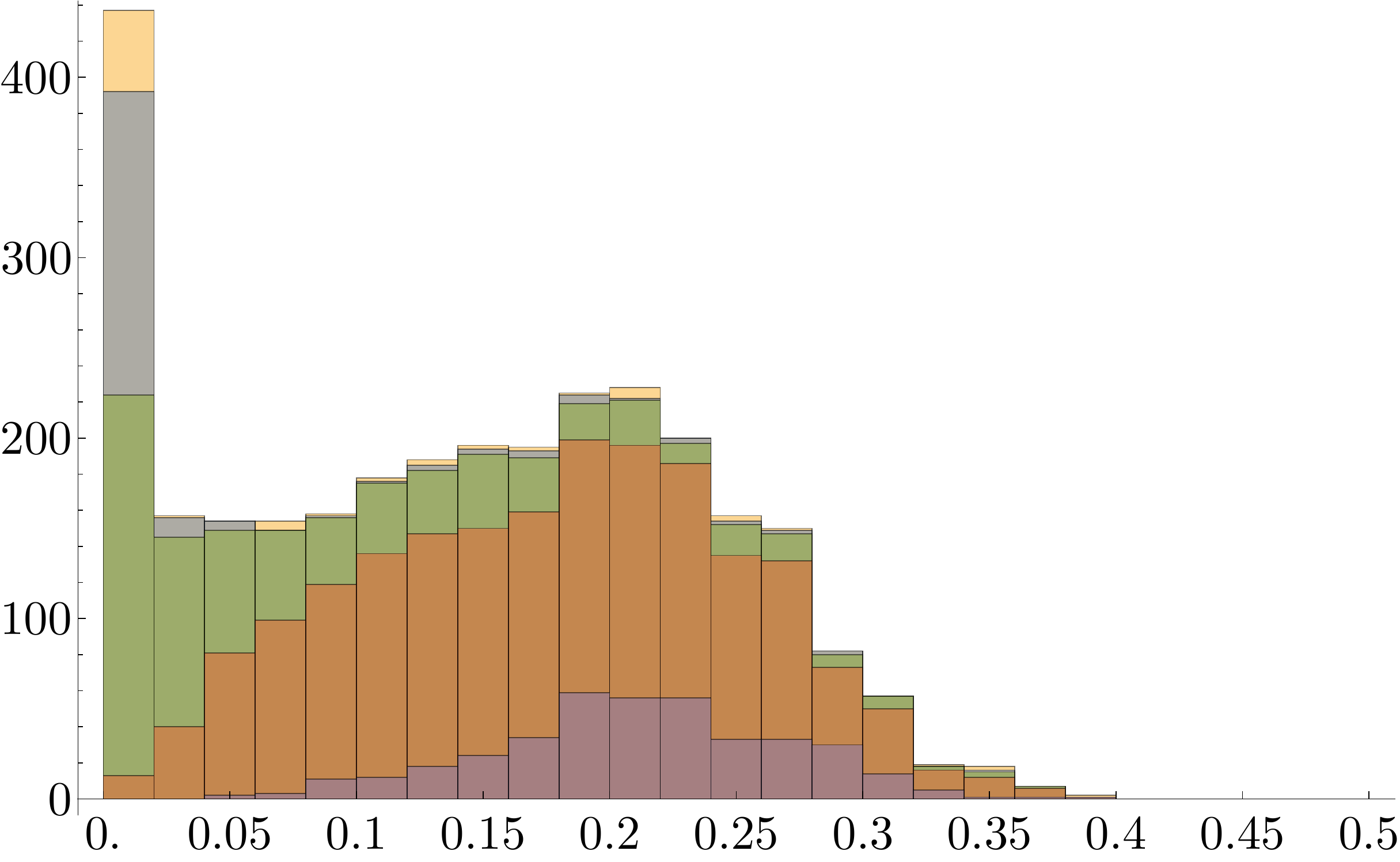}
			};
			\node [overlay] (x-axis) at (4, -2.8) {\scalebox{1.2}{$p$}};
			\node [overlay] at (.2, 2) {\fbox{20 steps}};
		\end{tikzpicture}
	\end{minipage}
	\caption{
		Distribution of projection probabilities for randomly sampled states, computed with the numerical maximization described in~\cref{sec:numerical_fid_max}.
		Both plots show the probabilities associated to a set of 3000 target states sampled from the uniform Haar distribution, for 15 and 20 steps.
		The light orange (upper) histograms represent the total number of target states found to correspond to a given range of probability.
		Starting from these datasets, we progressively removed the states that were found to reproduce the target states with fidelity less than $1 - 10^{-t}$, for various values of the threshold $t$.
		Light orange, grey, green, dark orange and purple (from top to bottom) histograms correspond to thresholds of respectively $t = 0, 2, 5, 10, 12$ (higher thresholds correspond to only a handful of states and are therefore omitted).
		This data further suggests a connection between the instability of the found solutions with respect to perturbations of the coin parameters, and the projection probability, as already hinted in \cref{fig:stabilities_5steps}:
		it is harder to find numerically with very good fidelity solutions corresponding to low projection probabilities because of their more unstable nature.
		It is also important to note that the solutions shown here are generally not the optimal ones,
		as the optimization algorithm only seeks to optimize the final fidelity,
		regardless of the corresponding projection probability.
	}
	\label{fig:prob_histograms_nmaximize}
\end{figure*}

The results of this approach are reported in \cref{fig:prob_histograms_nmaximize}, where we randomly generate a sample of target states, and find through numerical optimization the set of coin parameters and projections generating them.
It is worth noting that in this procedure we fixed the maximum number of iterations allowed for the maximization, not the precision with which the final fidelities are to be found.
This is done for the sake of efficiency, as some solutions are found to be more numerically unstable and hard to obtain with very high precisions through numerical optimization.
As a consequence, as reported in \cref{fig:prob_histograms_nmaximize}, some of the solutions are achieved with relatively low fidelities.
\Cref{fig:prob_histograms_nmaximize} also hints at a correlation between the more numerically unstable solutions and low projection probabilities:
almost all of the solutions that were reached with non-optimal fidelities (that is, fidelity less than 0.99) were also found to correspond to low projection probabilities.
This is consistent with the intuition provided by \cref{fig:stabilities_5steps},
that the lower probability solutions are more unstable with respect to variations of the coin parameters.
\Cref{fig:prob_histograms_nmaximize} shows that in $\sim$90\% (85\%) of the sampled instances we obtain strategies to generate, after 15 (20) steps, states with $p > 0.02$ and $\mathcal F > 0.99$.
It is worth noting that --- given the sharp difference between minimum and maximum probabilities reported in \cref{fig:minmax_probabilities}, and the above reasoning substantiated by \cref{fig:fid_vs_eps_varying_d,fig:prob_histograms_nmaximize} ---
there are strong reasons to believe that almost all target states can be achieved with both high fidelities and probabilities,
by properly fine-tuning the optimization algorithm.

\begin{figure*}[ht!]
\centering
\includegraphics[width=0.99\textwidth]{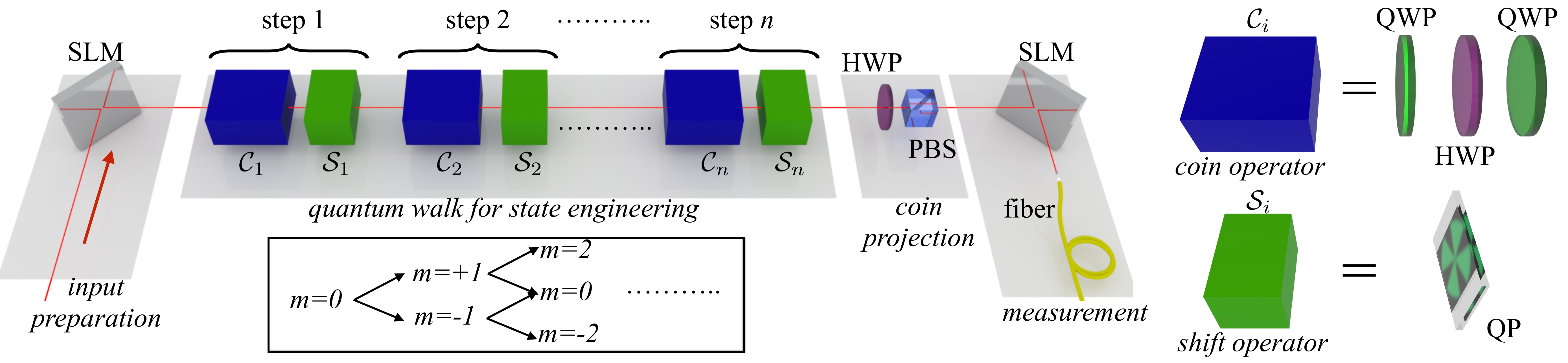}
\caption{
	Scheme for the proposed implementation of the quantum walk state engineering protocol using Orbital Angular Momentum (OAM) and Spin Angular Momentum of a photon.
	The input state is prepared in a superposition of $n$ modes with a Spatial Light Modulator (SLM).
	At each step, the coin operation is then realised in polarization with a sequence of a Quarter Wave Plate (QWP), a Half Wave Plate (HWP) and a second QWP, arranged to implement an arbitrary transformation.
	The shift operation in OAM space is implemented with a q-plate (QP), which shifts the OAM of $\pm 2q$ conditionally to the polarization state of the photon.
	For $q=1/2$, the shift is equal to $\pm 1$, with the corresponding evolution is  schematically shown in the lower box.
	Finally, the coin is projected in the $\vert + \rangle$ state by means of a HWP and a polarizing beam splitter (PBS).
	A second SLM followed by a single-mode fiber performs the measurement of the output state.
}
\label{fig:proposal_exp}
\end{figure*}
\section{Linear optical experimental implementation} The state engineering protocol here proposed can be tested with a DTQW in \ac{OAM} and \ac{SAM} of a photon.
The main feature of this scheme is that the evolution of the system lies in a single optical path~\cite{zhang2010implementation}, allowing for an implementation using a single photon.
The complete scheme is shown in Fig.~\ref{fig:proposal_exp}.
Position and coin degrees of freedom are encoded respectively in \ac{OAM} and \ac{SAM} of the photon,
with the walker's site described by the quantum number $m$ $\in$ $\mathbb{Z}$, eigenvalue of the \ac{OAM} along the propagation axis, and the coin state is encoded in left $\ket L$ and right $\ket R$ polarization states of the photon.
With this encoding, arbitrary coin operations are implemented propagating the photon through quarter (QWP) and half (HWP) wave plates.
The controlled shift operations are instead realized with q-plates (QP):
a birefringent liquid-crystal medium that rises and lowers the value of $m$ conditionally to the polarization state~\cite{marrucci2006optical},
changing the wavefront transverse profile of the photon without deflections.
More specifically, an input photon injected into a QP changes its state as $\vert m,R \rangle \rightarrow \vert m-2q,L \rangle$ and $\vert m,L \rangle \rightarrow \vert m+2q,R \rangle$, where $q$ is the topological charge of the QP.
Measurement of the coin in the $\vert + \rangle$ state is performed with a HWP and a polarizing beam splitter (PBS),
while the OAM state can be analysed with a spatial light modulator (SLM)~\cite{cardano2015quantum}.
Therefore, the DTQW is made up of consecutive optical units composed of wave plates (coin operators) and a QP (shift operators).
The number of optical elements scales polynomially with the number of steps, making this scheme scalable.
Finally, a SLM can be also employed before the quantum walk architecture to prepare the initial walker state spanning $n$ sites.

Among other possible architectures to implement the quantum state engineering are intrinsically-stable bulk interferometric schemes~\cite{broome2010discrete,kitagawa2012observation,vitelli2013joining}.
In this approach, the coin is implemented in the polarization degree of freedom,
while the shift operator is performed by introducing a spatial displacement of the optical mode conditionally to the polarization state of the photon.
Other approaches include integrated linear interferometers~\cite{sansoni2012twoparticle, crespi2013anderson, harris2015bosonic, pitsios2016photonic}, and fiber-loops architectures~\cite{schreiber2010photons, schreiber2012a, boutari2016large}. As mentioned, besides purely optical settings, one could envisage to adapt the present protocol to other physical systems that can host quantum walks \cite{manouchehri2014physical}, such as trapped atoms \cite{karski2009quantum} and ions \cite{schmitz2009quantum, zhringer2010realization} as well as cold atoms in lattices \cite{weitenberg2011singlespin, fukuhara2013microscopic, preiss2015strongly}.

\section{Conclusions} We provided a set of equations characterizing the states reachable by a coined quantum walk evolution, when letting the coin operation change from step to step.
We then derived a set of conditions characterizing the quantum states, spanning only the walker's positions, that are probabilistically reachable after projection of the coin at the end of the walk.
Finally, we proposed a protocol to experimentally implement the quantum state engineering scheme with linear optics, using orbital and spin angular momentum of a photon to encode spatial and coin degrees of freedom of the walker.
Given the ubiquity of quantum walks, our approach should facilitate the engineering of high-dimensional quantum states in a wide range of physical systems.

\begin{acknowledgments}
LI is grateful to Fondazione Angelo della Riccia for support. We acknowledge support from the EU project TherMiQ (grant agreement 618074), the ERC-Starting Grant 3DQUEST (3D-Quantum Integrated Optical Simulation; grant agreement no. 307783), the ERC-Advanced grant PHOSPhOR (Photonics of Spin-Orbit Optical Phenomena; grant agreement no. 694683),
 the UK EPSRC (grant EP/N508664/1), the Northern Ireland DfE, and the SFI-DfE Investigator Programme (grant 15/IA/2864). 
\end{acknowledgments}

\makeatletter\onecolumngrid@push\makeatother

\clearpage
\makeatletter\onecolumngrid@pop\makeatother

\bibliographystyle{apsrev4-1}
\bibliography{QW_bibliography}

\clearpage
\appendix
\section{Generalised reachability conditions}
\label{app:reachability_conditions}
We here describe a generalised form of the reachability results of~\cref{sec:reachability}.
In particular, we study the constraints associated to a state $\ket\Phi$ spanning $m+1$ sites, which is the output of at least $n$ steps of quantum walk evolution.
Note that while in~\cref{sec:reachability} we focused on the common $m=n$ case, we here consider the more general scenario with $m \ge n$.
Under these assumptions, $\ket\Phi$ is written as: 
\begin{equation}
	\ket\Phi =
	\mathcal W_{\mathcal C_{1}} \mathcal W_{\mathcal C_{2}} ... \mathcal W_{\mathcal C_{n}}
	\ket{\Phi_{in}},
	\label{eq:starting_point}
\end{equation}
for some set of coin operators $\{\mathcal C_i \}$ and initial state $\ket{\Phi_{in}}$.
We are here not imposing constraints on the form of $\ket{\Phi_{in}}$, which in particular does not have to satisfy any reachability condition of its own.
We want to show that, for any set of coin operators $\{\mathcal C_i \}$, \cref{eq:starting_point} implies that the amplitudes of $\ket\Phi$ satisfy the following set of equations:
\begin{gather}
	u_{1,\downarrow} = u_{m+1, \uparrow} = 0,
	\label{eq:vanishing_amplitudes}
	\\
	\sum_{i=1}^s \bs v_i^\dagger \bs v_{m-s+i} = 0,
	\text{ for } s=1,..., n-1,
	\label{eq:orthogonality_conditions_for_vs}
\end{gather}
where $\bs v_{i}$ is defined as
\begin{equation*}
	\bs v_i =
	\begin{pmatrix}
		u_{i,\uparrow} \\ u_{i+1, \downarrow}
	\end{pmatrix}.
\end{equation*}
The cases $n=2$ and $n=3$ were explicitly computed in~\cref{sec:reachability},
so let us assume the statement to be true for $n$ and show that this implies it for $n+1$.
The main idea is to see how each one of the equations in \cref{eq:orthogonality_conditions_for_vs} transforms after one step.
Let us denote the amplitudes after one step with $u'_{i,\alpha}$.
The relation between primed and unprimed amplitudes is therefore given by
\begin{equation*}
	\bs v_i^\prime
	= \mathcal C
	\begin{pmatrix}
		u_{i,\uparrow} \\ u_{i, \downarrow}
	\end{pmatrix},
	\text{ for every }
	i = 1, ..., m,
\end{equation*}
for some unitary coin operator $\mathcal C$,
with $\bs v'_i \equiv (u_{i,\uparrow}',u_{i+1,\downarrow}')$.
This directly implies that for any $i, j$,
\begin{equation}
	\begin{pmatrix}
		u_{i, \uparrow}^* & u_{i, \downarrow}^*
	\end{pmatrix}
	\begin{pmatrix}
		u_{j, \uparrow} \\ u_{j, \downarrow}
	\end{pmatrix}
	=
	\bs v^{\prime\dagger}_{i} \cdot \bs v^\prime_j.
	\label{eq:transition_us_to_vprimes}
\end{equation}
Consider then the $s$-th term in \cref{eq:orthogonality_conditions_for_vs}:
\begin{equation*}
	\bs v_1^\dagger \bs v_{m-s+1}
	+ \bs v_2^\dagger \bs v_{m-s+2}
	+ ...
	+ \bs v_{s-1}^\dagger \bs v_{m-1}
	+ \bs v_s^\dagger \bs v_m = 0.
\end{equation*}
Rewriting the LHS of the equation above in terms of the amplitudes, rearranging the terms, and remembering that $u_{1, \downarrow} = u_{m+1, \uparrow} = 0$, we get
{\footnotesize\begin{equation*}
\begin{aligned}
	\begin{pmatrix}
		u_{1, \uparrow}^* & u_{1, \downarrow}^*
	\end{pmatrix}
	\begin{pmatrix}
		u_{m-s+1, \uparrow} \\ u_{m-s+1, \downarrow}
	\end{pmatrix}
	+
	\begin{pmatrix}
		u_{2, \uparrow}^* & u_{2, \downarrow}^*
	\end{pmatrix}
	\begin{pmatrix}
		u_{m-s+2, \uparrow} \\ u_{m-s+2, \downarrow}
	\end{pmatrix} \\
	+ ... + 
	\begin{pmatrix}
		u_{s, \uparrow}^* & u_{s, \downarrow}^*
	\end{pmatrix}
	\begin{pmatrix}
		u_{m, \uparrow} \\ u_{m, \downarrow}
	\end{pmatrix}
	+
	\begin{pmatrix}
		u_{s+1, \uparrow}^* & u_{s+1, \downarrow}^*
	\end{pmatrix}
	\begin{pmatrix}
		u_{m+1, \uparrow} \\ u_{m+1, \downarrow}
	\end{pmatrix}.
\end{aligned}
\end{equation*}}
Using \cref{eq:transition_us_to_vprimes} the above becomes
\begin{equation*}
	\bs v^{\prime\dagger}_1 \bs v^\prime_{m-s+1} +
	\bs v^{\prime\dagger}_2 \bs v^\prime_{m-s+2} +
	... +
	\bs v^{\prime\dagger}_s \bs v^\prime_m +
	\bs v^{\prime\dagger}_{s+1} \bs v^\prime_{m+1},
\end{equation*}
or, equivalently,
$\displaystyle
	\sum_{i=1}^{s+1} \bs v^{\prime\dagger}_i \bs v^\prime_{m-s+i}.
$
This proves that
$
	\mathcal W_{\mathcal C}\mathcal W_{\mathcal C_1} \cdots \mathcal W_{\mathcal C_{n}}
	\ket{\Phi_{in}}
$
satisfies the set of $n-1$ constraints:
\begin{equation}
	\sum_{i=1}^{s}
	\bs v^{\prime\dagger}_i
	\bs v^\prime_{(m+1)-s+i} = 0,
	\text{ for }
	s = 2,...,n.
	\label{eq:general_orthogonality_almost_finished}
\end{equation}
To complete the proof, we only miss to show that \cref{eq:general_orthogonality_almost_finished} also holds for $s=1$, that is, that
$\bs v^{\prime\dagger}_1 \bs v^\prime_m = 0$.
But this follows immediately from the vanishing amplitudes of $\ket\Phi$ at first and $m$-th site, as already shown in~\cref{sec:2step_reachability}.

\section{Analytical solutions for two steps}
\label{app:analytical_sol_2steps}
We will in this section study the solutions obtained in the simplest case of 2 steps, focusing on what states are reachable, with what probabilities, and the stability of these solutions with respect to small perturbations of the coin parameters.

\begin{figure}[tb]
	\centering
	\includegraphics[width=\linewidth]{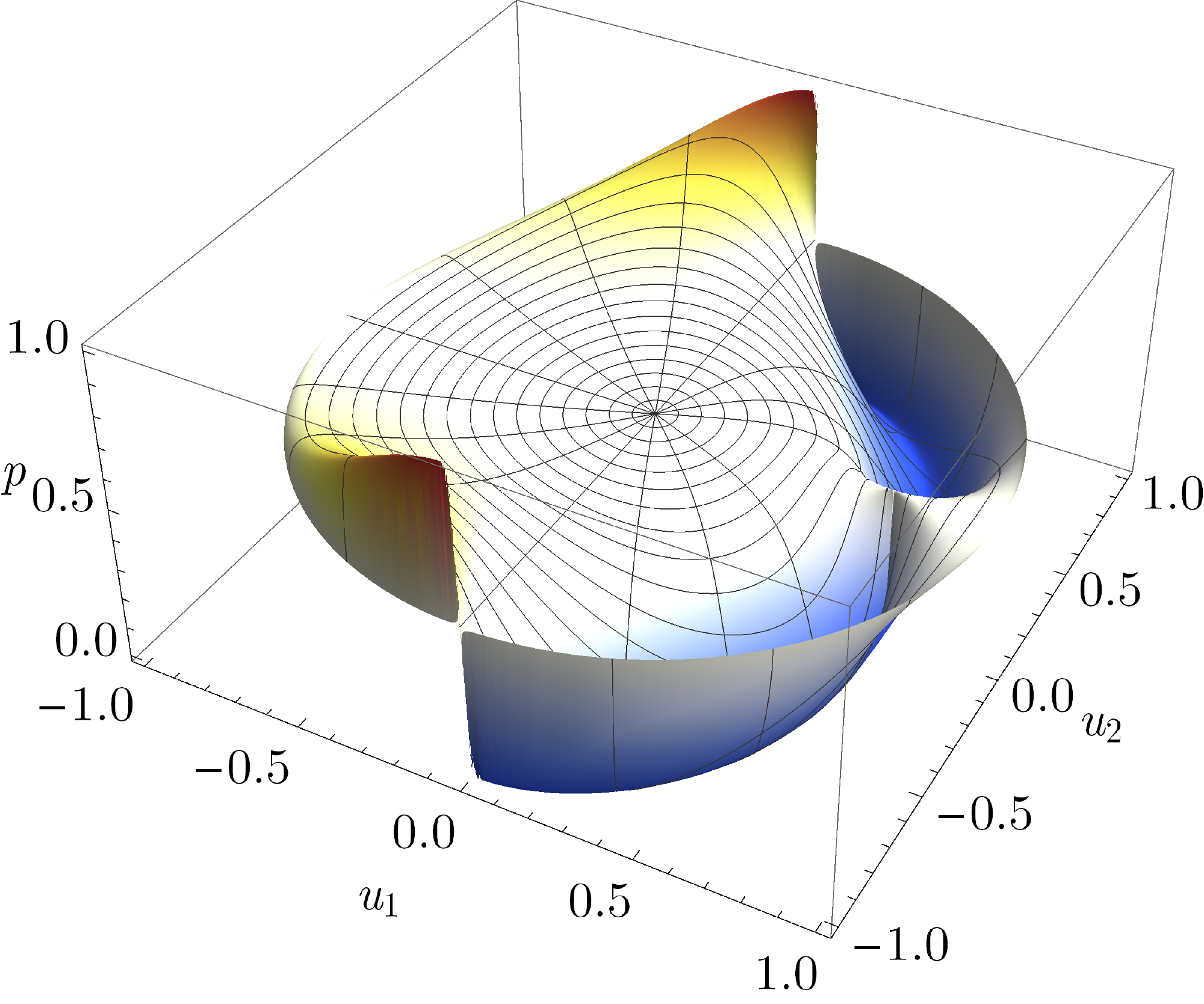}
	\caption{
		Probability given by \cref{eq:proj_prob_2steps} plotted against $u_1$ and $u_2$, for the special case of $u_1, u_2 \in \mathbb R$.
	}
	\label{fig:proj_prob_landscape_plot3d}
\end{figure}

\begin{figure}[tb]
	\centering
	\begin{tikzpicture}
		\node (img) {\includegraphics[width=\columnwidth]%
			{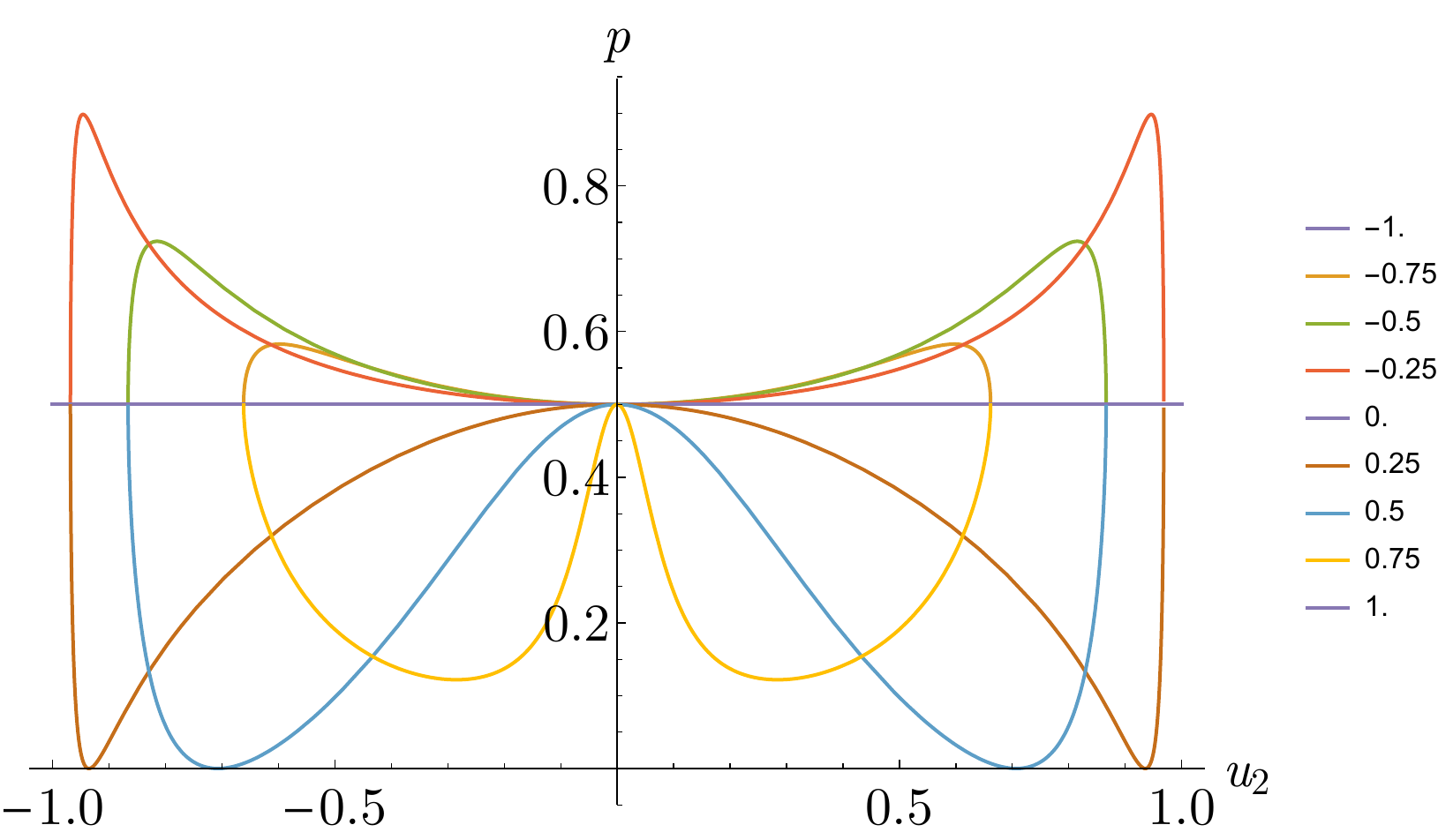}
		};
	\end{tikzpicture}
	\caption{
		Probability given by~\cref{eq:proj_prob_2steps} plotted against $u_2$, for various choices of $u_1$.
		Each line corresponds to a slice taken from \cref{fig:proj_prob_landscape_plot3d}.}
	\label{fig:proj_prob_landscape_slices}
\end{figure}

As already shown in~\cref{sec:focusing_walker_states}, for 2 steps the reachability conditions become
\begin{equation}
	u_1^* (u_2 - d_2) + d_2^* u_3 = 0,
	\label{eq:reachability_condition_2steps}
\end{equation}
which in the $\lvert u_1 \rvert \neq \lvert u_3 \rvert$ case has solution:
\begin{equation}
	d_2 = \frac{
		u_1(u_1^* u_2 + u_2^* u_3)
	}{
		\lvert u_1 \rvert^2 - \lvert u_3 \rvert^2
	}.
	\label{app:eq:solution_for_d2}
\end{equation}
Using the above expression we can compute the projection probability of reaching a specific target state.
For example, in the special case of $u_1, u_2, u_3 \in \mathbb R$, $u_3 \ge 0$, this probability is
\begin{equation}
\begin{aligned}
	p &= \frac{1}{2} \left[ u_1^2 + (u_2 - d_2)^2 + d_2^2 + u_3^2 \right]^{-1} \\
	&= \frac{
		\left( u_1 - u_3 \right)^2
	}{
		2(1 - u_2^2)(1 - 2 u_1 u_3)
	}
\end{aligned}
\label{eq:proj_prob_2steps}
\end{equation}
with $u_3 = \sqrt{1 - u_1^2 - u_2^2}$.
\Cref{fig:proj_prob_landscape_plot3d,fig:proj_prob_landscape_slices} show how this probability varies with $u_1$ and $u_2$.
Already in this simple case some interesting features emerge. For example, $p$ vanishes for $u_1 = u_3$.
On the other hand, for $u_1 = -u_3$, the probability does not vanish, which may be puzzling because in this case \cref{app:eq:solution_for_d2} itself is singular.
A more careful analysis of \cref{app:eq:solution_for_d2} shows however that $u_1 = -u_3$ corresponds to a removable singularity, consistently with the corresponding non-vanishing projection probability.

\begin{figure}[tb]
	\centering
	\includegraphics[width=\linewidth]{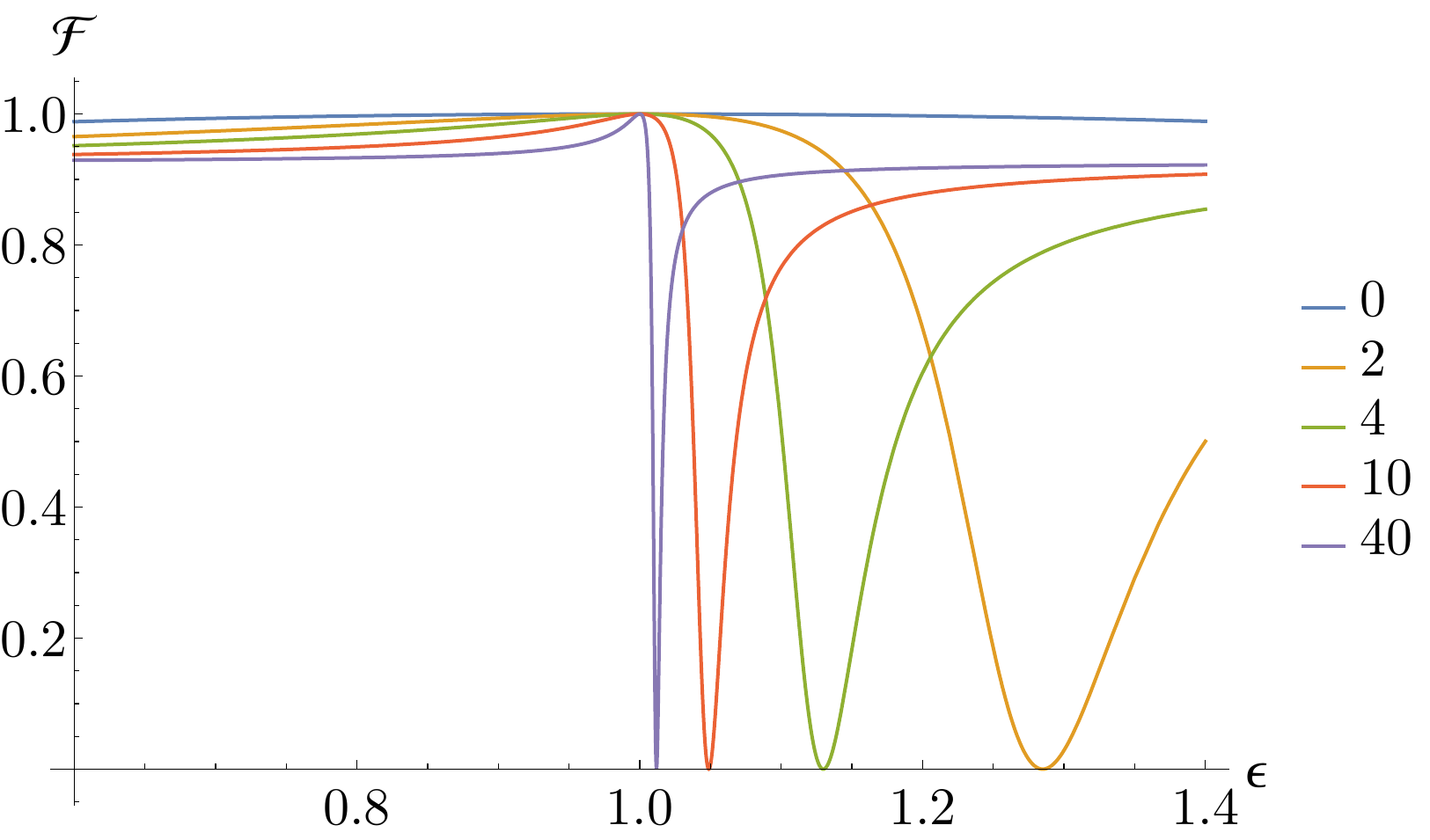}
	\caption{
		Fidelity vs a relative change of the value of a coin parameter, for different values of $d_I$.
		Starting from \cref{eq:state_degenerate_case}, we use as target state the normalized vector
		$\bs u = N (0.5 i, 0.2, 0.5 i)$, and for various values of $d_I$ we compute the coin parameters generating the full state shown in \cref{eq:state_degenerate_case}.
		We then take a single coin parameter, the parameter $\theta$ in the first step, and substitute it with $\epsilon \theta$, plotting the resulting fidelity between target and generated states as a function of $\epsilon$.
		Larger values of $d_I$ clearly correspond to higher instability.
	}
	\label{fig:fid_vs_eps_varying_d}
\end{figure}

When the $\lvert u_1 \rvert \neq \lvert u_3 \rvert$ condition is not met, the solution space changes significantly.
We here consider for example the case $u_1 = u_3 \neq 0$. Substituting this into \cref{eq:reachability_condition_2steps} we get the conditions
\begin{equation*}
	\begin{dcases}
		u_{1R}u_{2R} + u_{1I} u_{2I} = 0, \\
		u_{1R} (2d_I - u_{2I})
		+ u_{1I} (-2 d_R + u_{2R}) = 0,
	\end{dcases}
\end{equation*}
where $u_{iR}$ and $u_{iI}$ denote the real and imaginary parts of $u_i$, respectively.
A possible class of solutions of the above, obtained in the case $u_{1R} = 0$, is
\begin{equation}
	\scalemath{0.86}{
	\begin{pmatrix}
		i u_{1I} & 0 \\
		\frac{u_{2R}}{2}-i d_I & \frac{u_{2R}}{2} + i d_I \\
		0 & i u_{1I}
	\end{pmatrix}},\,\,
	u_{1R} = u_{2I} = 0,\,\, d_I \in \mathbb R.
	\label{eq:state_degenerate_case}
\end{equation}
Apart from the constraints we have to impose on the target $\bs u$ (implying that not all targets are reachable when these degenerate conditions are met), it is interesting to note that there is here an infinity of possible solutions, as $d_I$ can have any value.
However, these solutions correspond to different projection probabilities, which in the above case can be computed to be
\begin{equation*}
	p = \frac{1}{2\left( 2 u_{1I}^2 + u_{2R}^2/2 + 2 d_I^2 \right)},
\end{equation*}
the value of which ranges from 0, when $d_I \to \infty$, to a maximum of $1/2(2u_{1I}^2 + u_{2R}^2 / 2)$ when $d_I = 0$.
These different solutions also present different degrees of stability with respect to small changes of the coin parameters.
To illustrate this, in~\cref{fig:fid_vs_eps_varying_d} is shown how the fidelity varies when perturbing one of the coin parameters generating a state of the form \cref{eq:state_degenerate_case}, for various values of $d_I$.


\end{document}